\documentclass[pdflatex,sn-mathphys]{sn-jnl}

\usepackage{graphicx}%
\usepackage{multirow}%
\usepackage{amsmath,amssymb,amsfonts}%
\usepackage{amsthm}%
\usepackage{mathrsfs}%
\usepackage[title]{appendix}%
\usepackage{xcolor}%
\usepackage{textcomp}%
\usepackage{manyfoot}%
\usepackage{booktabs}%
\usepackage{algorithm}%
\usepackage{algorithmicx}%
\usepackage{algpseudocode}%
\usepackage{listings}%
\usepackage[sort&compress,numbers]{natbib}

\usepackage{amsmath, amssymb, amscd, amsthm, amsfonts, graphicx, hyperref,array, xcolor, geometry, float, scalerel, booktabs, multirow, appendix,setspace, bm, empheq, lipsum,arydshln, placeins, anyfontsize, enumitem}

\jyear{2021}%

\theoremstyle{thmstyleone}%
\theoremstyle{thmstyletwo}%

\theoremstyle{thmstylethree}%

\begin{document}

\title[A spatially resolved and lipid-structured model for early atherosclerosis]{
\begin{center}
    A spatially resolved and lipid-structured model for macrophage populations in early human atherosclerotic lesions
\end{center}
}



\author*[1,4]{\fnm{Keith L.} \sur{Chambers}}\email{keith.chambers@maths.ox.ac.uk}

\author[2]{\fnm{Mary R.} \sur{Myerscough}}\email{mary.myerscough@sydney.edu.au}

\author[3]{\fnm{Michael G.} \sur{Watson}}\email{michael.watson1@unsw.edu.au}

\author[1,4]{\fnm{Helen M.} \sur{Byrne}}\email{helen.byrne@maths.ox.ac.uk \clearpage}

\affil*[1]{\orgdiv{Wolfson Centre for Mathematical Biology}, \orgname{Mathematical Institute, University of Oxford}, \orgaddress{\street{Andrew Wiles Building, Radcliffe Observatory Quarter, Woodstock Road}, \city{Oxford}, \postcode{OX2 6GG}, \state{Oxfordshire}, \country{United Kingdom}}}

\affil[2]{\orgdiv{School of Mathematics and Statistics}, \orgname{University of Sydney}, \orgaddress{\street{Carslaw Building, Eastern Avenue, Camperdown}, \city{Sydney}, \postcode{2006}, \state{New South Wales}, \country{Australia}}}

\affil[3]{\orgdiv{School of Mathematics and Statistics}, \orgname{University of New South Wales}, \orgaddress{\street{The Red Centre, University Mall, UNSW, Kensington}, \city{Sydney}, \postcode{2052}, \state{New South Wales}, \country{Australia}}}

\affil[4]{\orgdiv{Ludwig Institute for Cancer Research}, \orgname{University of Oxford}, \orgaddress{\street{Old Road Campus Research Build, Roosevelt Dr, Headington}, \city{Oxford}, \postcode{OX3 7DQ}, \state{Oxfordshire}, \country{United Kingdom}}}

\abstract{Atherosclerosis is a chronic inflammatory disease of the artery wall. The early stages of atherosclerosis are driven by interactions between lipids and monocyte-derived-macrophages (MDMs). The mechanisms that govern the spatial distribution of lipids and MDMs in the lesion remain poorly understood. In this paper, we develop a spatially-resolved and lipid-structured model for early atherosclerosis. The model development and analysis are guided by images of human coronary lesions by Nakashima \textit{et al.} \cite{nakashima2007early}. Consistent with their findings, the model predicts that lipid initially accumulates deep in the intima due to a spatially non-uniform LDL retention capacity. The model also qualitatively reproduces the global internal maxima in the Nakashima images only when the MDM mobility is sufficiently sensitive to lipid content, and MDM lifespan sufficiently insensitive. Introducing lipid content-dependence to MDM mobility and mean lifespan produced minimal impact on model behaviour at early times, but strongly impacted lesion composition at steady state. Increases to the sensitivity of MDM lifespan to lipid content yield lesions with fewer MDMs, less total lesion lipid content and reduced mean MDM infiltration depth. Increases to the sensitivity of MDM mobility to lipid content also reduces the MDM infiltration depth, but increases the proportion of lipid-laden MDMs. We find that MDM lipid content increases with spatial depth, regardless of blood LDL and HDL content. These results shed light on the mechanisms that drive spatial variation in the composition of early atherosclerotic lesions, and the role of macrophage lipid content in disease progression.
}




\maketitle

\clearpage
\section{Introduction}\label{sec: Intro}

Atherosclerosis is a chronic inflammatory disease of the artery wall and a leading cause of death around the world \cite{lusis2000atherosclerosis}. It is triggered by the retention of low-density lipoprotein (LDL) particles in the \textit{(tunica) intima}, the innermost layer of artery wall \cite{boren2016central}. LDL particles enter the intima from the bloodstream and become trapped in the extracellular matrix. These retained LDL (rLDL) particles are rapidly modified via oxidation and aggregation \cite{oorni2000aggregation, oorni2021aggregation}. The accumulation of rLDL initiates a signal cascade that attracts monocyte-derived macrophages (MDMs) to the intima from the bloodstream. MDMs are the dominant immune cell type by number in early lesions \cite{willemsen2020macrophage}. They ingest extracellular lipids such as LDL and offload lipid onto high-density lipoprotein (HDL) particles, which also enter from the bloodstream \cite{kloc2020role}. The sustained influx of lipid and the death of lipid-laden MDMs may cause the lesion to develop into an atherosclerotic plaque with a large lipid core \cite{guyton1996development, gonzalez2017macrophage}. Such plaques are vulnerable to rupture, which allows the thrombotic content of the lipid core to enter the circulation and cause a blood clot.
Plaque rupture is the most common cause of myocardial infarction and a leading cause of ischaemic strokes \cite{costopoulos2017plaque, rothwell2007atherothrombosis}. The mechanisms that govern the spatial distribution of lipids and MDMs in the lesion remain poorly understood.

Spatially resolved data on lipid and MDM accumulation in early human lesions is scarce. To the best of our knowledge, the only study  which presents such data is by Nakashima \textit{et al.} \cite{nakashima2007early}. Spatially resolved images of lipid and MDM densities in early human coronary lesions collected by Nakashima \textit{et al.} are shown in Fig. \ref{fig: Nakashima}. 
%
%
Inspection of the lipid densities in Fig. \ref{fig: Nakashima}(b,e,h) shows that lipid first accumulates deep in the intima, towards the media. Nakashima \textit{et al.} note that biglycan, an extracellular matrix proteoglycan, occurs in the same spatial location as the early distribution of lipids, suggesting that the early deposition profile is due to a non-uniform LDL retention capacity. The images in Fig. \ref{fig: Nakashima}(k, n, q) show a second wave of lipid influx that coincides with the influx of MDMs (c.f. (l, o, r)).
{\color{black} Importantly, the MDM density remains localised to the intima, with negligible numbers of macrophages near the media. More recent work by Nakagawa \textit{et al. } \cite{nakagawa2021accumulation} further supports the hypothesis that MDMs do not traverse into the media from the intima in early human lesions. By the `Pathological intimal thickening (PIT) with foam cells' stage (see (q) and (r)), localised accumulation of lipid-laden MDMs drives the lipid and macrophage densities to peak within the intima. Indeed, the region of greatest lipid density approximately coincides with that of greatest MDM density. We regard this 
behaviour as a key feature that our mathematical model should reproduce under 
pathological conditions.
%
%


\begin{figure}[h]
    \centering
    \includegraphics[width=0.9\textwidth]{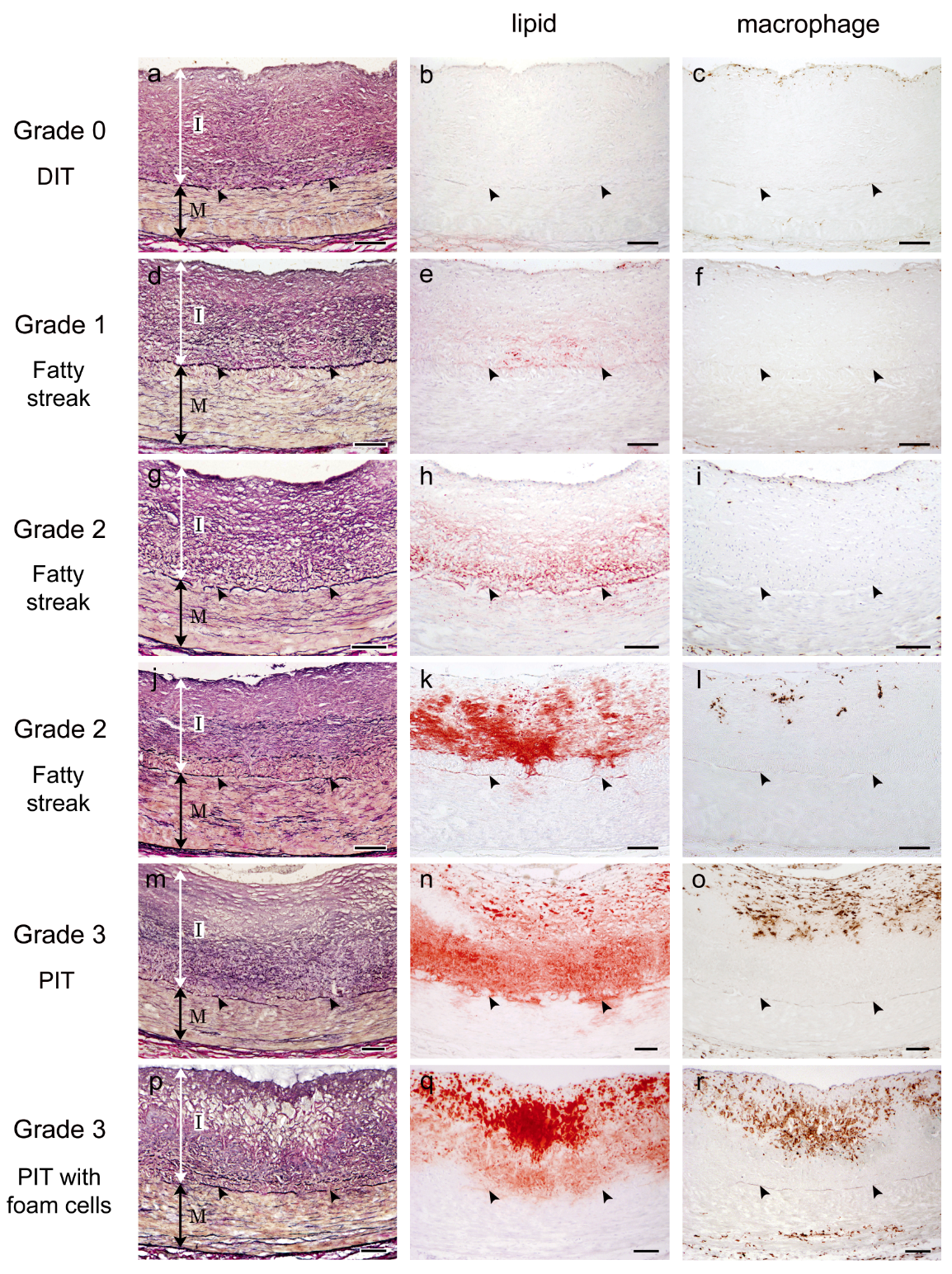}
    \caption{\textbf{Early stages of human coronary atherosclerosis. } The images show representative lipid and macrophage densities across six stages of lipid accumulation, ranging from diffuse intimal thickening (DIT), fatty streaks, to pathological intimal thickening (PIT). The endothelium is at the top of each image, and corresponds to $x = 0$ in our model. The internal elastic lamina (IEL) which separates the tunica intima (labeled `I') from the tunica media (labeled `M') is marked with black arrows, and corresponds to $x = X$. Note how lipid first accumulates deep in the tunica intima. The lipid and macrophage densities eventually both peak within the tunica intima, in plots (q) and (r) respectively. Bars on the bottom-right represent $0.1$mm. Reproduced, with permission, from  Nakashima \textit{et al.} \cite{nakashima2007early}
    .}
    \label{fig: Nakashima}
\end{figure}

Of particular relevance {\color{black} to our model} is the impact of lipid ingestion on MDM behaviour. We note first that lipid-laden MDMs exhibit a decreased lifespan relative to unladen MDMs. This is due to the cytotoxicity of free cholesterol, which promotes controlled cell death (apoptosis) \cite{yao2001free}. Lipid-laden MDMs are also less mobile than unladen MDMs, due in part to cytoskeletal changes \cite{pataki1992endocytosis, chen2019lysophosphatidic}. Both effects are likely to impact the spatial distribution of MDMs (and consequently, lipid) since the distribution of MDM lipid content is unlikely to be spatially uniform. The impact of these lipid content-dependent effects on the MDM and lipid spatial distributions is presently unknown, and another focus of the present study. 

Mathematical modelling of atherosclerosis is a growing field of study \cite{parton2016computational, avgerinos2019mathematical,cai2021mathematical, mc2022modeling}. The literature includes models of LDL infiltration \cite{prosi2005mathematical, yang2006modeling, yang2008low}, plaque mechanics \cite{fok2012mathematical, watson2020multiphase} and lesion immunology. Models of lesion immunology include agent-based simulations \cite{corti2020fully, bayani2020spatial}, ODE models \cite{bulelzai2012long, cohen2014athero, islam2015mathematical, thon2018quantitative, lui2021modelling, xie2022well} and spatial PDE models \cite{calvez2009mathematical,fok2012growth, hao2014ldl, chalmers2015bifurcation, mukherjee2019reaction, mohammad2020integrated, ahmed2023macrophage}. Importantly, with the notable exceptions of Fok \cite{fok2012growth} and Mirzaei \textit{et al.} \cite{mohammad2020integrated}, the existing spatial PDE models have not been compared to images of plaque formation. 
Moreover, existing spatial models adopt a simplistic treatment of MDM lipid content by distinguishing between lipid-poor MDMs, simply called `macrophages', and lipid-laden MDMs, called `foam cells'. Since MDM lipid accumulation is gradual, such an approach makes it difficult to analyse the impact of lipid content-dependent MDM lifespan and mobility on lesion development. 

Several recent studies have used structured population models to describe gradual lipid accumulation in MDMs \cite{ford2019efferocytosis, watson2023lipid, chambers2024lipid, chambers2024blood}. Of particular relevance to the present study is the model of Chambers \textit{et al.} \cite{chambers2023new}, which we extend here to account for 
heterogeneity in both MDM lipid content and spatial position.
Our work contributes to the growing literature on spatially-resolved structured population models \cite{celora2023spatio, fiandaca2022phenotype, pan2022propagation, boulouz2022spatially, hu2019spatial, liu2015hopf}. 
In existing models, spatial movement is represented by including diffusion terms for unbiased motion and advection terms for directed motion. Movement may be independent of structure, as in \cite{pan2022propagation}, or dependent on the structure variable (e.g., Fiandaca \textit{et al.} consider phenotype-dependent cell speeds \cite{fiandaca2022phenotype}). In the current work, we assume for simplicity that MDM motion is unbiased but dependent on lipid content. {\color{black} We further neglect local proliferation of macrophages in our model, justified by observations that proliferation is limited in aortic macrophage populations \cite{williams2020limited}, and may only be prominent in late rather than early lesions \cite{robbins2013local}. Hence, we draw no distinction between `macrophages' and `MDMs' in the present work. An example of how proliferation may be incorporated in a lipid-structured framework is given in \cite{chambers2024lipid}.} We use our continuum model to address the following key questions:\\
\begin{enumerate}
     \item[Q1.] {\color{black} \textbf{When does our model reproduce the key qualitative features of Fig. \ref{fig: Nakashima}?} In particular, the initial accumulation of lipid deep in the intima, towards the tunica media, and the eventual formation of internal global maxima in both the lipid and MDM densities.
     }
     \\
     \item[Q2.] {\color{black} \textbf{How do lipid content-dependent MDM lifespan and mobility impact lesion composition?}} \\
     \item[Q3.] {\color{black} \textbf{Is MDM lipid content correlated with spatial depth?}} \\
\end{enumerate} 

The remainder of this paper is organised as follows. Sect. \ref{sec: model development} explains the model development. Sect. \ref{sec: results} contains the results{\color{black}, which consist primarily of numerical solutions and steady state analysis of the MDM density}. We consider the model in increasing complexity by introducing lipid content-dependent lifespan and mobility one at a time, and then simultaneously. Finally, in Sect. \ref{sec: discussion} we discuss the results and how they address Q1-Q3. 

\FloatBarrier
  
\section{Model development} \label{sec: model development}

In this section we present a spatially-resolved model describing the dynamics of lipids and MDMs in early atherosclerosis. The MDM population is further structured according to lipid content. A model schematic is given in Fig. \ref{fig: schematic}.

\begin{figure}
    \centering
    \includegraphics[width=0.99\linewidth]{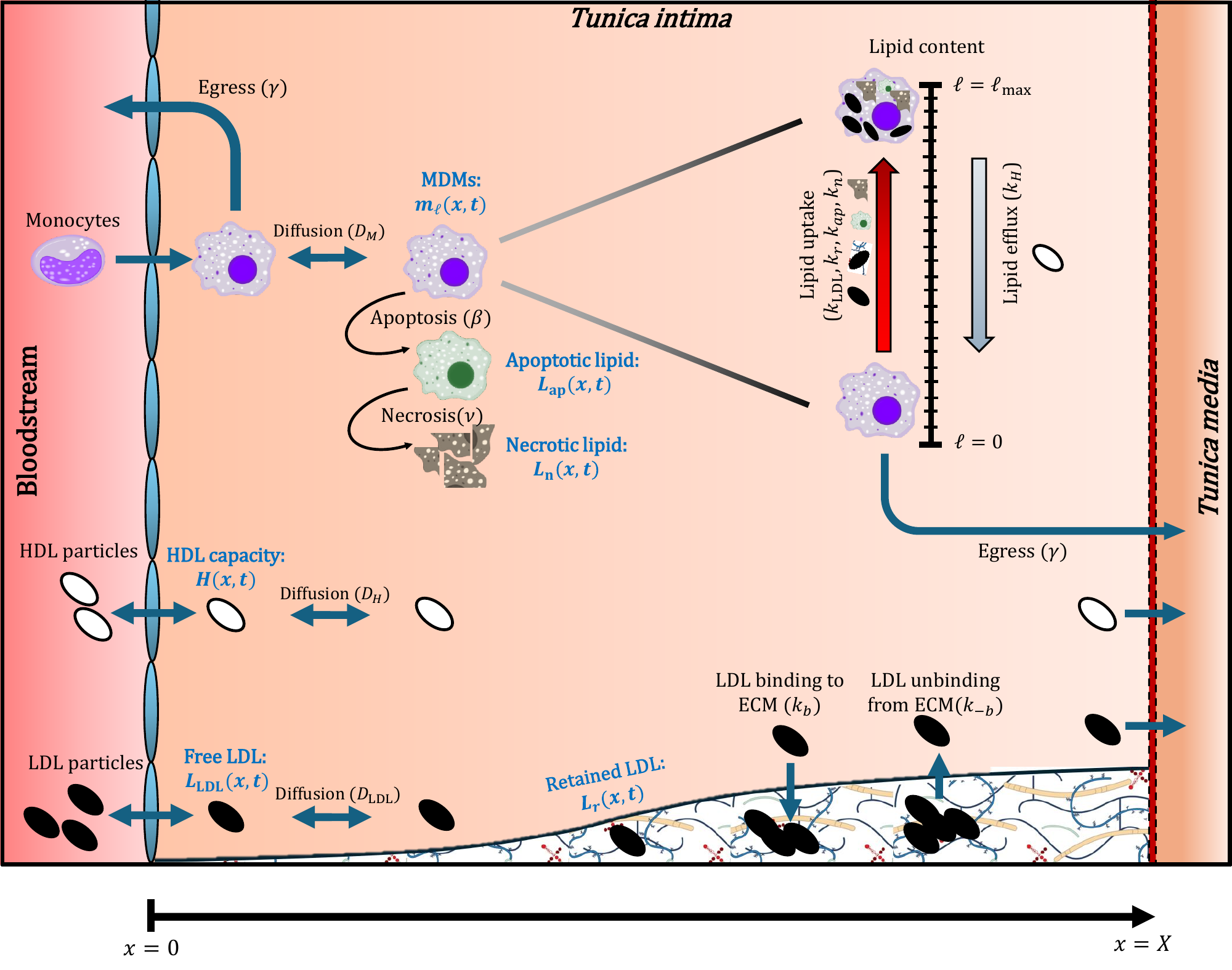}
    \caption{\textbf{Schematic depiction of the model. } The spatial domain is the \textit{tunica intima}, which is bordered by the bloodstream at the endothelium, $x = 0$, and the \textit{tunica media} at the IEL, $x = X$. The dependent variables are highlighted in blue. The discrete lipid-structured MDM population is illustrated in the top-right; MDMs increase their $\ell$-index via lipid uptake, and decrease their $\ell$-index via lipid efflux to HDL. We incorporate a non-uniform LDL retention capacity due to abundant LDL-retaining proteoglycans towards the \textit{tunica media}. We omit the smaller diffusivities of apoptotic and necrotic lipid, and the lipid-dependence of MDM mobility and apoptosis {\color{black} in this schematic} for legibility.}
    \label{fig: schematic}
\end{figure}

\subsection{Assumptions and definitions}

We formulate the model as a system of partial differential equations, with associated boundary and initial conditions. The equations are defined in one spatial dimension, $0 \leq x \leq X$, which represents distance from the endothelium ($x=0$) to the IEL ($x=X$). This Cartesian formulation assumes that the domain is sufficiently far from the edges of the lesion that net circumferential and axial motions are negligible. We assume for simplicity that the domain length is fixed. This approximation is reasonable for early lesions but is unlikely to hold for late-stage plaques. 

We further assume that MDMs take up and offload lipid in finite increments of mass $\Delta a > 0$, and structure the MDM population according to lipid content. Specifically, we let $m_{\ell}(x,t) \geq 0$ denote the density of MDMs with lipid content $a_0 + \ell \Delta a$ at position $x$ and time $t \geq 0$. The index $\ell$ runs over non-negative values: $\ell = 0, 1, \dots, \ell_\text{max}$, so that MDM lipid content ranges from $a_0 \geq 0$, the endogenous content, to $a_0 + \kappa$, where $\kappa = \ell_\text{max} \Delta a$ is the maximum amount of lipid that an MDM can ingest.

We introduce additional dependent variables to describe the extracellular environment. We denote by $L_{\text{\scaleto{LDL}{3.5pt}}}(x,t) \geq 0$, $L_{\text{r}}(x,t) \geq 0$, $L_{\text{ap}}(x,t) \geq 0$ and $L_{\text{n}}(x,t) \geq 0$ the mass densities of free LDL lipid, retained LDL lipid, apoptotic lipid and necrotic lipid respectively. We let $H(x,t)$ be the lipid capacity of HDL particles in the lesion. Finally, we denote by $S(x,t) \geq 0$ the mass density of inflammatory mediators. 

\subsection{Model equations}

\noindent \textbf{MDMs.} We propose that the MDM population evolves according to the following PDEs:
\begin{align}
    \begin{split} \label{eqn: ml}
        \frac{\partial m_\ell}{\partial t} &= \underbrace{D_M \bigg[ 1 - \psi_D \Big( \frac{\ell}{\ell_\text{max}} \Big) \bigg] \frac{\partial^2 m_\ell}{\partial x^2}}_{\text{diffusion}} - \underbrace{\frac{\beta m_\ell}{1 - \psi_\beta \big( \frac{\ell}{\ell_\text{max}} \big)}}_{\text{apoptosis}} \\
        &\quad + \underbrace{\bm{k_L} \cdot \bm{L} \, \, \big[ (\ell_\text{max}-\ell + 1) m_{\ell-1} - (\ell_\text{max} - \ell)m_{\ell} \big]}_{\text{lipid uptake}} \\
    &\quad + \underbrace{k_H H \, \, \, \, \, \big[ (\ell + 1) m_{\ell + 1} - \ell m_{\ell} \big]}_{\text{lipid efflux to HDL}}, \\
    \end{split}
\end{align}
for all  $0 \leq x \leq 1$ and $t \geq 0$. Eqs. \eqref{eqn: ml} hold for $\ell = 0, 1, \dots, \ell_\text{max}$, subject to the closure conditions:
\begin{align}
    m_{-1} = m_{\ell_\text{max} + 1} = 0,
\end{align}
which ensure that MDMs with unphysical lipid loads do not arise in the model.

The first term on the RHS of Eqs. \eqref{eqn: ml} describes MDM transport by random motion.
Motivated by the observation that macrophage mobility decreases upon conversion to foam cells \cite{pataki1992endocytosis}, we assume that MDM mobility decreases linearly with $\ell$ so that MDMs with low lipid loads are more mobile than lipid-laden cells. The parameter $0 \leq \psi_D \leq 1$ determines how sensitive MDM diffusion is to lipid load. If $\psi_D = 0$ then there is no lipid-dependence and all MDMs have diffusion coefficient $D_M$. When $\psi_D > 0$, MDM diffusivity decreases from $D_M$ at $\ell = 0$ to $D_M(1 - \psi_D)$ at $\ell = \ell_\text{max}$. When $\psi_D = 1$, fully lipid laden MDMs are immobile. 

The second term on the RHS of Eqs. \eqref{eqn: ml} represents MDM apoptosis. We suppose that the apoptosis rate is an increasing function of $\ell$, so that lipid-laden macrophages die more frequently than those with lower lipid loads, due to the cytotoxicity of intracellular free cholesterol \cite{yao2001free}. More specifically, we assume that the mean MDM lifespan decreases linearly with $\ell$, from $\beta^{-1}$ when $\ell = 0$ to $\beta^{-1} (1 - \psi_\beta)$ when $\ell = \ell_\text{max}$. The parameter $0 \leq \psi_\beta \leq 1$ determines the sensitivity of mean lifespan on lipid content; if $\psi_\beta = 0$ then all MDMs share a common mean lifespan, $\beta^{-1}$, regardless of their lipid content, whereas if $\psi_\beta = 1$ then the MDM mean lifespan decreases to zero at $\ell = \ell_\text{max}$ (MDMs at capacity die instantly, giving $m_{\ell_\text{max}} = 0$). Our formulation assumes that the apoptosis rate increases with $\ell$. 
This nonlinear dependence is reasonable since lipid handling is a normal macrophage function and lipid-induced apoptosis is a response to \textit{excess} lipid accumulation \cite{remmerie2018macrophages}. Therefore, increases to lower lipid loads are likely to have less impact on macrophage viability than increases to lipid-laden cells. 

{\color{black} The third term on the RHS of Eqs. \eqref{eqn: ml} accounts for lipid uptake. Following \cite{chambers2023new}, we assume that MDMs take up lipid in increments of mass $\Delta a$ at a rate proportional to the product of their available capacity and the extracellular lipid density: $k_i L_i (\ell_\text{max} - \ell) \geq 0$. Here $i \in \{ \text{\scaleto{LDL}{5pt}}, \text{r}, \text{ap}, \text{n} \}$ specifies the type of lipid ingested. For notational brevity in Eqs.\eqref{eqn: ml}, we introduce vectors for the uptake rates, $\bm{k_L}:= ( k_{\text{\scaleto{LDL}{3.5pt}}},  k_{\text{r}}, k_\text{ap}, k_\text{n})$, and the extracellular lipids, $\bm{L} := ( L_{\text{\scaleto{LDL}{3.5pt}}},  L_{\text{r}}, L_\text{ap}, L_\text{n})$, so that $\bm{k_L}\cdot \bm{L} = k_{\text{\scaleto{LDL}{3.5pt}}} L_{\text{\scaleto{LDL}{3.5pt}}} + k_{\text{r}}L_{\text{r}} + k_\text{ap}L_\text{ap} + k_\text{n}L_\text{n}$.

The final term on the RHS of Eqs. \eqref{eqn: ml} accounts for lipid efflux to HDL. Again, following \cite{chambers2023new}, we assume that MDMs offload lipid to HDL particles in increments of mass $\Delta a$ at a rate proportional to the product of their non-endogenous lipid content and the HDL lipid capacity in the lesion: $k_H H \ell \geq 0$.}


We close Eqs. \eqref{eqn: ml} by imposing the following boundary conditions for $\ell = 0, 1, \dots, \ell_\text{max}$:
\begin{align} \label{eqn: ml bconds}
    -D_M \frac{\partial m_\ell}{\partial x} &= 
    \begin{cases}
         R(t) \delta_{\ell, 0} - \gamma m_{\ell} & \text{at } x=0; \\
         \gamma m_{\ell} & \text{at } $x=X$,
    \end{cases}
\end{align}
where $\delta_{\ell, 0}$ is the Kronecker-Delta. Eqs. \eqref{eqn: ml bconds} assume that newly recruited MDMs enter the lesion at $x = 0$ with only their endogenous lipid content at rate $R(t)$. The recruitment rate is a first-order Hill function of the inflammatory mediator density at the endothelium:
\begin{align}
    R(t) := \frac{\sigma_M S(0,t)}{S_{\text{c50}} + S(0,t)},
\end{align}
with a maximal recruitment rate of $\sigma_M$ and half-maximal recruitment occurring when $S(0,t) = S_{\text{c50}}$. Eqs. \eqref{eqn: ml bconds} also account for MDM egress through either boundary. More specifically, we assume that MDMs at $x = 0, X$ exit the lesion at a rate proportional to their mobility: $\gamma \big[1 - \psi_D \big( \frac{\ell}{\ell_\text{max}} \big)\big]$. Since internal elastic lamina egress rates have not been measured, to the best of our knowledge, we assume for simplicity that egress rates at $x = 0$ and $x = X$ are equal. We state initial conditions for $m_\ell$ together with the other variables later in this section (see Eq. \eqref{eqn: init}).
\\

\noindent \textbf{Extracellular lipids. } We assume the densities of free and retained LDL evolve according to the PDEs:
\begin{align}
    \begin{split} \label{eqn: LDL}
        \frac{\partial L_{\text{\scaleto{LDL}{3.5pt}}}}{\partial t} &= \underbrace{D_{\text{\scaleto{LDL}{3.5pt}}} \frac{\partial^2 L_{\text{\scaleto{LDL}{3.5pt}}}}{\partial x^2}}_{\text{diffusion}}  \quad - \underbrace{k_b L_{\text{\scaleto{LDL}{3.5pt}}} [K_{\text{r}}(x) - L_{\text{r}}]}_{\text{retention via ECM binding}} + \underbrace{k_{-b} L_{\text{r}}}_{\text{unbinding}} \\
        &\quad - \underbrace{k_{\text{\scaleto{LDL}{3.5pt}}} \Delta a L_{\text{\scaleto{LDL}{3.5pt}}} \sum_{\ell=0}^{\ell_\text{max}} (\ell_\text{max} - \ell) m_{\ell}}_{\text{uptake by MDMs}},
    \end{split} \\
    \begin{split} \label{eqn: Lr}
        \frac{\partial L_{\text{r}}}{\partial t} &=   k_b L_{\text{\scaleto{LDL}{3.5pt}}} [K_{\text{r}}(x) - L_{\text{r}}] - k_{-b} L_{\text{r}} - k_{\text{r}} \Delta a L_{\text{r}} \sum_{\ell=0}^{\ell_\text{max}} (\ell_\text{max} - \ell) m_{\ell}.
    \end{split}
\end{align}
The first {\color{black} RHS} term in Eq. \eqref{eqn: LDL} accounts for random motion of free LDL, with diffusivity $D_{\text{\scaleto{LDL}{3.5pt}}} > 0$. We also assume that free LDL binds to ECM proteoglycans with rate constant $k_b$ in a capacity-limited manner, and unbinds at rate $k_{-b}$. Motivated by Fig. \ref{fig: Nakashima}, we suppose that the capacity for LDL retention $K_r(x)$  is a sigmoidal function of position:
\begin{align} 
    K_\text{r}(x) := \Bar{K}_r \cdot \bigg( \frac{2}{1 + e^{\theta (1/2-x/X)}} \bigg), \quad  \text{where } \Bar{K}_r = \frac{1}{X} \int_0^X K_r(x) dx.\label{eqn: K_r}
\end{align}
Here $\Bar{K}_r$ is the mean rLDL capacity in the intima and $\theta > 0$ determines the steepness of $K_\text{r}(x)$ (plots of the functional form for  $K_r(x)/\Bar{K}_r$ are shown in Fig. \ref{fig: Kr(x)}). The final terms in Eqs. \eqref{eqn: LDL}-\eqref{eqn: Lr} account for MDM uptake.

\begin{figure}
    \centering
    \includegraphics[width=0.6\textwidth]{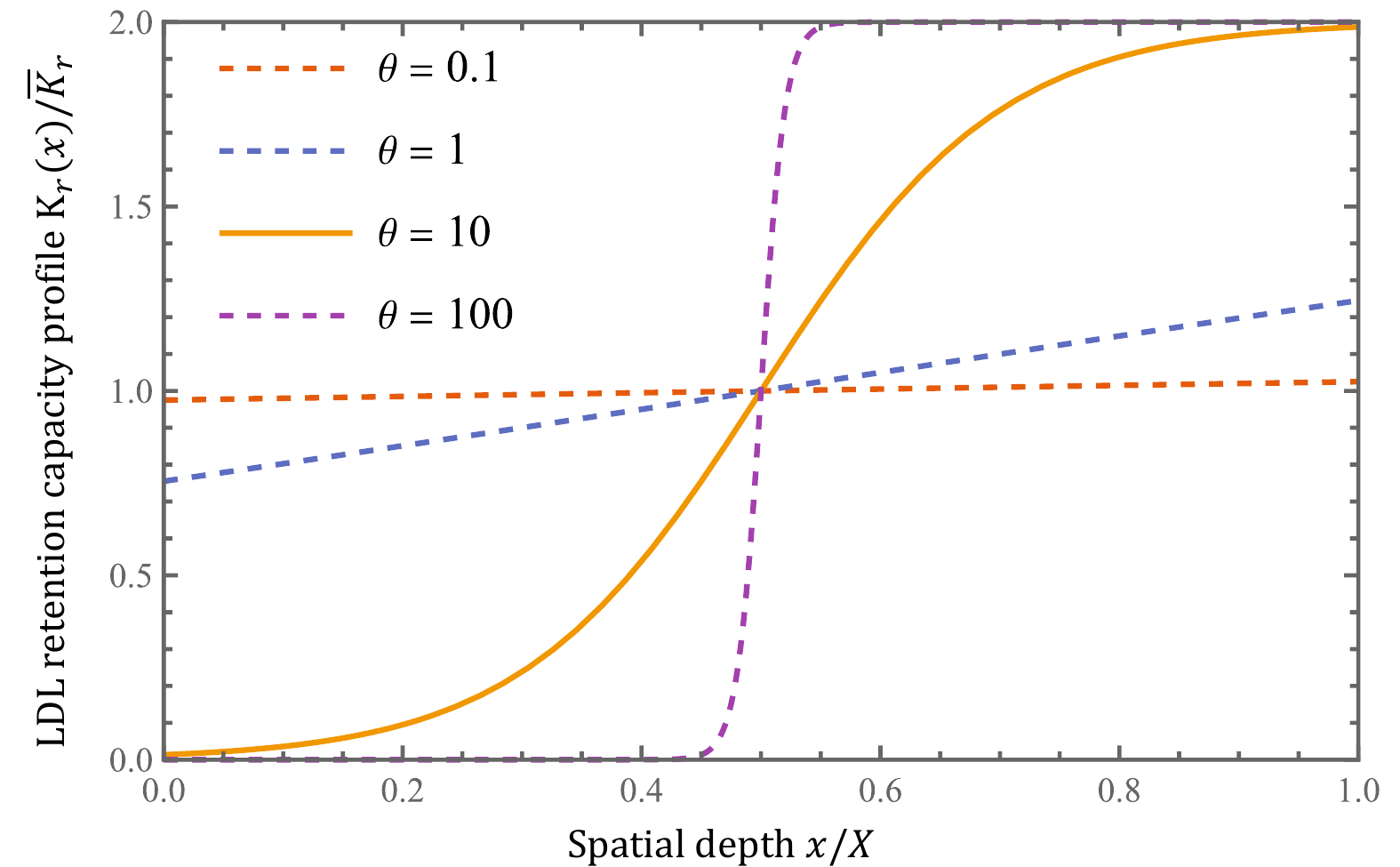}
    \caption{\textbf{Spatial dependence of the LDL retention capacity, defined in Eq. \eqref{eqn: K_r}. } Higher values of $\theta$ increase the sharpness of the step. We choose $\theta = 10$ for our numerical simulations to mimic the smooth increase in LDL retention with depth seen in Fig. \ref{fig: Nakashima}(e,h).}
    \label{fig: Kr(x)}
\end{figure}

We close Eq. \eqref{eqn: LDL} by imposing the following boundary conditions:
\begin{align}
    -D_{\text{\scaleto{LDL}{3.5pt}}} \frac{\partial L_{\text{\scaleto{LDL}{3.5pt}}}}{\partial x} &= 
    \begin{cases}
        P_L^{(0)} (L^\star - L_{\text{\scaleto{LDL}{3.5pt}}}) & \text{at } x = 0; \\
        P_L^{(1)} (L_{\text{\scaleto{LDL}{3.5pt}}} - 0) & \text{at } x = X.
    \end{cases} \label{eqn: LDLbconds}
\end{align}
Eqs. \eqref{eqn: LDLbconds} account for the exchange of free LDL between the tunica intima and the lumen/tunica media. Parameters $P_L^{(0)}$ and $P_L^{(1)}$ denote the permeability to free LDL of the endothelium and internal elastic lamina, respectively. We assume that the density of LDL lipid in the blood is given by a constant $L^\star$. Motivated by human data reported by Smith \textit{et al.}~\cite{smith1982plasma}, we suppose further that the density of LDL lipid in the tunica media is negligible. 

We propose that the apoptotic and necrotic lipid densities satisfy:
\begin{align}
    \begin{split}
        \frac{\partial L_{\text{ap}}}{\partial t} &= \underbrace{D_\text{ap} \frac{\partial^2 L_\text{ap}}{\partial x^2}}_{\text{diffusion}} + \sum_{\ell=0}^{\ell_\text{max}}  \bigg[ \underbrace{\frac{\beta (a_0 + \ell \Delta a)}{1 - \psi_\beta (\frac{\ell}{\ell_\text{max}})}}_{\text{apoptosis}} - \underbrace{k_\text{ap} \Delta a L_\text{ap} (\ell_\text{max} - \ell)}_{\text{uptake by MDMs}} \bigg] m_\ell \\
        &\quad - \underbrace{\nu L_\text{ap}}_{\text{necrosis}},
    \end{split} \label{eqn: Lap} \\
    \frac{\partial L_\text{n}}{\partial t} &=  D_\text{n} \frac{\partial^2 L_\text{n}}{\partial x^2 } + \nu L_{\text{ap}} - k_\text{n} \Delta a L_\text{n} \sum_{\ell=0}^{\ell_\text{max}} (\ell_\text{max} - \ell) m_{\ell}. \label{eqn: Ln}
\end{align}
The terms on the first line of the RHS of Eq. \eqref{eqn: Lap} account for diffusion, lipid deposition due to MDM apoptosis and uptake by MDMs. We further assume that apoptotic cells undergo secondary necrosis at rate $\nu$, which provides linear sink and source terms in Eqs. \eqref{eqn: Lap} and \eqref{eqn: Ln}, respectively. 

We close Eqs. \eqref{eqn: Lap}-\eqref{eqn: Ln} by imposing the following boundary conditions:
\begin{align}
    \frac{\partial L_\text{ap}}{\partial x} = \frac{\partial L_\text{n}}{\partial x} = 0 \quad  \text{at } x = 0, X. \label{eqn: Lap Ln bconds}
\end{align}
Eqs. \eqref{eqn: Lap Ln bconds} assume for simplicity that the endothelium and internal elastic lamina are impermeable to both apoptotic and necrotic lipid.
\\


\noindent \textbf{HDL lipid capacity. } We assume that the lipid capacity of the HDL particles is governed by the following PDE:
\begin{align} \label{eqn: H}
        \frac{\partial H}{\partial t} &= \underbrace{D_H \frac{\partial^2 H}{\partial x^2}}_{\text{diffusion}} \quad - \underbrace{k_H \Delta a H \sum_{\ell = 0}^{\ell_\text{max}} \ell m_{\ell}}_{\text{lipid efflux by MDMs}}.
\end{align}
Eq. \eqref{eqn: H} accounts for HDL diffusion and MDM lipid efflux to HDL.

We close Eq. \eqref{eqn: H} by imposing the following boundary conditions:
\begin{align}
    -D_H \frac{\partial H}{\partial x} &= 
    \begin{cases}
        P_H^{(0)} (H^\star - H) & \text{at } x = 0; \\
        P_H^{(1)} (H - 0) & \text{at } x = X.
    \end{cases} \label{eqn: Hbconds}
\end{align}
Eqs. \eqref{eqn: Hbconds} describe the flux of HDL lipid capacity  between the tunica intima and the lumen/tunica media. The constants $P_H^{(0)}$ and $P_H^{(1)}$ denote the permeability to HDL of the endothelium and internal elastic lamina, respectively. We assume that the blood density of HDL lipid capacity is a constant $H^\star$. Similar to our treatment of LDL lipid, we assume that the HDL lipid capacity density in the tunica media is negligible. \\

\noindent \textbf{Inflammatory mediators. } Finally, we propose that the density of inflammatory mediators $S$ evolves according to the following PDE:
\begin{align}
    \frac{\partial S}{\partial t} &= \underbrace{D_S \frac{\partial^2 S}{\partial x^2}}_{\text{diffusion}} + \underbrace{\alpha L_\text{r}}_{\substack{\text{resident response} \\ \text{to LDL retention}}} - \underbrace{\delta_S S}_\text{natural decay}. \label{eqn: S}
\end{align}
The first term on the RHS of Eq. \eqref{eqn: S} accounts for mediator diffusion, with diffusion constant $D_S$. Consistent with the response-to-retention hypothesis, we assume that inflammatory mediators are produced by resident cells at a rate proportional to the density of rLDL; the signal cascades which first stimulate MDM recruitment are, as yet, unknown, but are thought to be due to excessive LDL retention \cite{williams2005lipoprotein}. Finally, we assume that inflammatory mediators undergo natural decay at rate $\delta_S$. 

We close Eq. \eqref{eqn: S} with the boundary conditions:
\begin{align}
    -D_S \frac{\partial S}{\partial x} &= \begin{cases}
        P_S^{(0)} (0 - S) & \text{at } x = 0; \\
        P_S^{(1)} (S - 0) & \text{at } x = X.
    \end{cases}\label{eqn: Sbconds}
\end{align}
In Eqs. \eqref{eqn: Sbconds} we assume, for simplicity, that the mediator densities in the lumen and tunica media are negligible, and that mediators leave the lesion via the endothelium and internal elastic lamina with permeabilities $P_S^{(0)}$ and $P_S^{(1)}$ respectively. \\

\noindent \textbf{Initial conditions. } We impose the following conditions at $t = 0$:
\begin{align}
    &m_\ell = 0 \quad \forall \ell = 0, 1, \dots, \ell_\text{max}, & &L_{\text{\scaleto{LDL}{3.5pt}}} = L_\text{r} = L_\text{ap} = L_\text{n} = H = S = 0. \label{eqn: init}
\end{align}
Eqs. \eqref{eqn: init} assert that the model lesion is initially devoid of MDMs, extracellular lipid, HDL and inflammatory mediators. 

\subsection{Total MDM density and lipid content}

It will be useful, when discussing our results, to define the total MDM density, $M(x,t)$, and MDM lipid content, $L_M(x,t)$: 
\begin{align} 
    &M(x,t) := \sum_{\ell = 0}^{\ell_\text{max}} m_\ell(x,t), & &L_M(x,t) := \sum_{\ell=0}^{\ell_\text{max}} (a_0 + \ell \Delta a) m_\ell(x,t). \label{eqn: M, L_M defs}
\end{align}
Differentiating Eqs. \eqref{eqn: M, L_M defs} with respect to $t$ and substituting Eqs. \eqref{eqn: ml} gives:
\begin{align}
    \frac{\partial M}{\partial t} &= D_M \frac{\partial^2}{\partial x^2} \bigg[ M - \psi_D \sum_{\ell = 0}^{\ell_\text{max}}  \frac{\ell m_\ell}{\ell_\text{max}}  \bigg] - \beta \bigg[ M + \psi_\beta \sum_{\ell = 0}^{\ell_\text{max}} \frac{\ell m_\ell}{\ell_\text{max} - \psi_\beta \ell } \bigg], \label{eqn: M} \\
    \begin{split}
        \frac{\partial L_M}{\partial t} &= D_M \frac{\partial^2}{\partial x^2} \bigg[ L_M - \psi_D \sum_{\ell = 0}^{\ell_\text{max}}  \frac{(a_0 + \ell \Delta a) \ell m_\ell}{\ell_\text{max}}  \bigg]  + \bm{k_L} \cdot \bm{L} [(\kappa + a_0) M - L_M] \\ &\qquad \quad - \beta \bigg[ L_M + \, \, \psi_\beta \sum_{\ell = 0}^{\ell_\text{max}} \frac{(a_0 + \ell \Delta a)\ell m_\ell}{\ell_\text{max}  - \psi_\beta \ell } \bigg]  - k_H H (L_M - a_0 M).
    \end{split} \label{eqn: L_M}
\end{align}
The RHS terms in Eq. \eqref{eqn: M} account for MDM diffusion and apoptosis. The RHS terms in Eq. \eqref{eqn: L_M} describe diffusion of MDM lipid, lipid uptake, apoptosis and lipid efflux respectively. We may also rewrite Eqs. \eqref{eqn: LDL}, \eqref{eqn: Lr}, \eqref{eqn: Lap}, \eqref{eqn: Ln}, \eqref{eqn: H} and \eqref{eqn: S} as:
\begin{align}
    \begin{split}
        \frac{\partial L_{\text{\scaleto{LDL}{3.5pt}}}}{\partial t} &= D_{\text{\scaleto{LDL}{3.5pt}}} \frac{\partial^2 L_{\text{\scaleto{LDL}{3.5pt}}}}{\partial x^2} - k_b L_{\text{\scaleto{LDL}{3.5pt}}} [K_\text{r}(x) - L_\text{r}] + k_{-b}L_r \\
        &\quad - k_{\text{\scaleto{LDL}{3.5pt}}} L_{\text{\scaleto{LDL}{3.5pt}}} [(\kappa + a_0) M - L_M],     
        \end{split}\\
    \begin{split}
        \frac{\partial L_\text{r}}{\partial t} &= k_b L_{\text{\scaleto{LDL}{3.5pt}}} [K_\text{r}(x) - L_\text{r}] - k_{-b}L_r  - k_\text{r} L_\text{r} [(\kappa + a_0) M - L_M],
    \end{split} \\
    \begin{split}
        \frac{\partial L_\text{ap}}{\partial t} &= D_\text{ap} \frac{\partial^2 L_\text{ap}}{\partial x^2} + \beta \bigg[ L_M +  \psi_\beta \sum_{\ell = 0}^{\ell_\text{max}} \frac{(a_0 + \ell \Delta a)\ell m_\ell}{\ell_\text{max}  - \psi_\beta \ell } \bigg] \\
        &\quad - \nu L_\text{ap} - k_\text{ap}L_\text{ap} [(\kappa + a_0) M - L_M],
    \end{split} \\
    \frac{\partial L_\text{n}}{\partial t} &= D_\text{n} \frac{\partial^2 L_\text{n}}{\partial x^2} +  \nu L_\text{ap} - k_\text{n} L_\text{n} [(\kappa + a_0) M - L_M], \\
    \frac{\partial H}{\partial t} &= D_H \frac{\partial^2 H}{\partial x^2} - k_H H (L_M - a_0 M), \\
    \frac{\partial S}{\partial t} &= D_S \frac{\partial^2 S}{\partial x^2} + \alpha L_\text{r} - \delta_S S. \label{eqn: S subsystem}
\end{align}

Eqs. \eqref{eqn: M}-\eqref{eqn: S subsystem} form a closed subsystem when MDM diffusion and apoptosis are lipid content-independent ($\psi_D = \psi_\beta = 0$); otherwise they are coupled to Eqs. \eqref{eqn: ml}. We derive boundary conditions for $M$ and $L_M$  by differentiating Eqs. \eqref{eqn: M, L_M defs} with respect to $x$ and substituting Eqs. \eqref{eqn: ml bconds}: 
\begin{align} 
    -D_M \frac{\partial M}{\partial x} &= 
    \begin{cases}
             R(t) - \gamma M & \text{at } x=0; \\
         \gamma M & \text{at } $x=X$,
    \end{cases} \label{eqn: M bconds}\\
        -D_M \frac{\partial L_M}{\partial x} &= 
    \begin{cases}
             a_0 R(t) - \gamma L_M & \text{at } x=0; \\
         \gamma L_M & \text{at } $x=X$.
    \end{cases} \label{eqn: L_M bconds}
\end{align}
The boundary conditions \eqref{eqn: LDLbconds}, \eqref{eqn: Lap Ln bconds}, \eqref{eqn: Hbconds} and \eqref{eqn: Sbconds} for the remaining dependent variables are unchanged. 

\subsection{Parameter values}

\setlength\dashlinedash{0.5pt}
\setlength\dashlinegap{1.5pt}
\setlength\arrayrulewidth{0.3pt}
\begin{table}[]
    \caption{Model parameters} 
    \centering
    \begin{tabular}{cp{4.5cm}lp{1cm}}  
    \toprule
    Parameter    & Interpretation & Estimate & Source \\
    \midrule
    $L^{\star}$ & Lumen LDL lipid density & $0-440$ mg/dL & \cite{lee2012characteristics, orlova1999three}  \\
    $H^{\star}$ & Lumen HDL lipid capacity & $0-230$ mg/dL & \cite{madsen2017extreme} \\ 
    $P_L^{(0)}$ & Endothelial permeability to LDL & $0.37$ mm month$^{-1}$ & \cite{nielsen1996transfer, holzapfel2005determination}\\
    $P_H^{(0)}$ & Endothelial permeability to HDL & $0.74$ mm month$^{-1}$ &  \cite{stender1981transfer} \\
    $P_L^{(1)}$ & IEL permeability to LDL & $1.1$ mm month$^{-1}$ & \cite{penn1994relative}\\
    $P_H^{(1)}$ & IEL permeability to HDL & $2.2$ mm month$^{-1}$ &  \cite{penn1994relative} \\
    $P_S^{(0)}$ & Endothelial permeability to mediators & $5.0$ mm month$^{-1}$ &  \cite{yuan2010effect} \\
    $P_S^{(1)}$ & IEL permeability to mediators & $15$ mm month$^{-1}$ &  \cite{penn1994relative} \\
    $\overline{K}_r$ & Mean LDL retention capacity & $1900$ mg/dL & \cite{wight2018role, liu2023co}  \\
    $k_b$ & LDL retention rate & $0.008$ dL/mg month$^{-1}$ & \cite{bancells2009high, smith1982plasma}  \\
    $k_{-b}$ & LDL unbinding rate & $1.8$ month$^{-1}$ & \cite{bancells2009high, smith1982plasma} \\
    $\sigma_M$ & Maximum MDM entry rate & $6800$ mm$^{-2}$ month$^{-1}$ & \cite{williams2009transmigration, lee2019sirt1} \\    
    $\beta$ & Unladen MDM apoptosis rate & $1.0$ month$^{-1}$ & \cite{yona2013fate, williams2020limited} \\
    $\gamma$ & Unladen MDM egress rate & $0.08$ mm month$^{-1}$ & \cite{williams2018limited, lee2019sirt1}\\
    $\nu$ & Secondary necrosis rate & $37$ month$^{-1}$ & \cite{saraste2000morphologic} \\
    $a_0$ & MDM endogenous lipid & $27$pg &  \cite{sokol1991changes, cooper2022cell}\\
    $\kappa$ & MDM lipid capacity & $800$pg &  \cite{ford2019efferocytosis} \\
    $k_{\text{\scaleto{LDL}{3.5pt}}}$ & LDL lipid uptake rate & $0.00034$ dL/mg month$^{-1}$ &  \cite{sanda2021aggregated} \\
    $k_{\text{r}}$ & rLDL lipid uptake rate & $0.023$ dL/mg month$^{-1}$ &  \cite{sanda2021aggregated} \\
    $k_{\text{ap}}$ & Apoptotic lipid uptake rate & $0.060$ dL/mg month$^{-1}$ &  \cite{taruc2018quantification, schrijvers2005phagocytosis} \\
    $k_{\text{n}}$ & Necrotic lipid uptake rate & $0.015$ dL/mg month$^{-1}$ &  \cite{brouckaert2004phagocytosis} \\
    $k_{H}$ & Lipid efflux rate & $0.34$ dL/mg month$^{-1}$ &  \cite{kritharides1998cholesterol, woudberg2018pharmacological} \\
    $S^{\text{c50}}$ & Mediator density for half-max. MDM influx & $5$ ng/mL &  \cite{o2015pro, pugin1993lipopolysaccharide} \\
    $\delta_S$ & Mediator natural decay rate  & $1600$ month$^{-1}$ &  \cite{liu2021cytokines} \\
    $\alpha$ & Resident inflammatory mediator production per rLDL lipid  & $3.5 \times 10^{-5}$ month$^{-1}$ & \cite{williams2020limited}   \\
    $\Delta a$ & MDM uptake/efflux increment of lipid & $16$ pg &  \cite{kontush2007preferential, taefehshokr2021rab} \\
    $\ell_\text{max}$ & MDM lipid capacity per $\Delta a$ & $100$ &  $= \kappa/\Delta a$ \\
    \\
    \hdashline \\  
    $X$ & Tunica intima width & $0.40$ mm  & \cite{nakashima2007early} \\
    $\theta$ & LDL retention profile shape & $10$ & \cite{nakashima2007early} \\
    $D_{M}$ & Diffusivity of unladen MDMs & $0.02$ mm$^{2}$ month$^{-1}$ & \cite{adlerz2016substrate} \\
    $D_{\text{ap}}$ & Diffusivity of apoptotic cells & $0.005$ mm$^{2}$ month$^{-1}$ & est. \\
    $D_{\text{n}}$ & Diffusivity of necrotic cells & $0.005$ mm$^{2}$ month$^{-1}$ & est. \\
    $D_{\text{\scaleto{LDL}{3.5pt}}}$ & Diffusivity of free LDL & $9.2$ mm$^{2}$ month$^{-1}$ & \cite{thon2019spatially} \\
    $D_{H}$ & Diffusivity of HDL & $22$ mm$^{2}$ month$^{-1}$ & \cite{thon2019spatially} \\
    $D_{S}$ & Diffusivity of mediators & $500$ mm$^{2}$ month$^{-1}$ & \cite{ross2018diffusion} \\
    $\psi_D$ & MDM mobility sensitivity to lipid load & $\in [0,1]$  & - \\
    $\psi_\beta$ & MDM lifespan sensitivity to lipid load & $\in [0,1]$  & - \\
    \bottomrule
    \end{tabular}
        \label{tab: parameters}
\end{table}

The model parameters are summarised in Table \ref{tab: parameters}. Parameters above the dotted lines are in common with, or have direct analogues to, the parameters in \cite{chambers2024blood}. In particular, we explore a range of plausible values for the lumen blood densities of LDL lipid, $L^{\star} = 0-440$mg/dL, and HDL lipid capacity, $H^{\star} = 0-230$mg/dL; their values depend on lifestyle choices (e.g., diet) and, as such, are modifiable in practice. The parameters below the dotted lines do not have analogues in \cite{chambers2024blood}. We briefly explain their estimates below.

Our estimate of the tunica intima width, $X \approx 0.40$mm, is derived from Fig. \ref{fig: Nakashima}. These images indicate that tunica intima width does not vary substantially {\color{black} between the Fatty Streak and early PIT stages, where our model is based}. {\color{black} We further estimate $\theta = 10$ from Fig. \ref{fig: Nakashima} so that the LDL retention capacity, $K_r({x})$,  increases smoothly with depth (see Eq. \eqref{eqn: K_r} and Fig. \ref{fig: Kr(x)})}. 

The diffusivity parameters vary substantially in magnitude. The value for lipid-unladen MDMs, $D_M = 0.014$ mm$^{2}$ month$^{-1}$, derives from data on macrophage migration paths in \cite{adlerz2016substrate}. We assume that the diffusivity of apoptotic and necrotic cells, $D_{\text{ap}}=D_\text{n}=0.003$ mm$^{2}$ month$^{-1}$, is smaller than that of MDMs since dead cells do not move as independent agents. The diffusivities of free LDL, $D_{\text{\scaleto{LDL}{3.5pt}}} = 9.2$mm$^{2}$ month$^{-1}$, HDL, $D_H = 22$mm$^{2}$ month$^{-1}$, and mediators, $D_S = 500$mm$^{2}$ month$^{-1}$, are substantially higher. 

\subsection{Non-dimensionalisation} \label{sec: nondim}
\begin{table}[]
    \caption{Dimensionless parameters in Eqs. \eqref{eqn: ml nondim}-\eqref{eqn: init nondim}}. 
    \centering
    \begin{tabular}{clp{6cm}c}  
    \toprule
    Parameter & Definition  & Interpretation & Estimate \\
    \midrule
    $\Tilde{L}^\star$ & $\frac{\beta }{a_0 \sigma_M} L^{\star}  $ & Lumen LDL lipid density & $0-10$ \\
    $\Tilde{H}^\star$ & $\frac{ \beta }{a_0 \sigma_M} H^{\star}$ & Lumen HDL lipid capacity & $0-5$ \\
    $\Tilde{{K}}_\text{r}$ & $\frac{\beta}{a_0 \sigma_M} \overline{K}_r $ & Mean LDL retention capacity & $25$ \\ 
    $\Tilde{P}_L^{(0)}$ & $\frac{1}{X \beta} P_{L}^{(0)}$ & Endothelial LDL permeability & $0.9$ \\
    $\Tilde{P}_H^{(0)}$ & $\frac{1}{X \beta} P_{H}^{(0)}$ & Endothelial HDL permeability & $1.8$ \\
    $\Tilde{P}_S^{(0)}$ & $\frac{1}{X \beta} P_{S}^{(0)}$ & Endothelial mediator permeability & $13$ \\
    $\Tilde{P}_L^{(1)}$ & $\frac{1}{X \beta} P_{L}^{(1)}$ & Internal elastic lamina LDL permeability & $2.7$ \\
    $\Tilde{P}_H^{(1)}$ & $\frac{1}{X \beta} P_{H}^{(1)}$ & Internal elastic lamina HDL permeability & $5.4$ \\
    $\Tilde{P}_S^{(1)}$ & $\frac{1}{X \beta} P_{S}^{(1)}$ & Internal elastic lamina mediator permeability & $38$ \\
    $\Tilde{k}_b$ & $\frac{a_0 \sigma_M}{\beta^2 X} k_b$ & LDL retention rate & $2.7$ \\
    $\Tilde{k}_{-b}$ & $\frac{1}{\beta} k_{-b}$ & LDL unbinding rate & $1.8$ \\
    $\Tilde{\gamma}$ & $\frac{1}{\beta} \gamma$ & MDM egress rate & $0.2$ \\
    $\Tilde{\nu}$ & $\frac{1}{\beta} \nu$ & Secondary necrosis rate & $37$ \\
    $\Tilde{\kappa}$ & $\frac{1}{a_0} \kappa$ & MDM lipid capacity per unit endogenous lipid & $29$ \\
    $\Tilde{k}_{\text{\scaleto{LDL}{3.5pt}}}$ & $\frac{a_0 \sigma_M}{\beta^2 X} k_{\text{\scaleto{LDL}{3.5pt}}}$ & LDL uptake rate & $0.016$ \\
    $\Tilde{k}_{\text{r}}$ & $\frac{a_0 \sigma_M}{\beta^2 X} k_{\text{r}}$ & rLDL uptake rate & $1.1$ \\
    $\Tilde{k}_{\text{ap}}$ & $\frac{a_0 \sigma_M}{\beta^2 X} k_{\text{ap}}$ & Apoptotic lipid uptake rate & $5.5$ \\
    $\Tilde{k}_{\text{n}}$ & $\frac{a_0 \sigma_M}{\beta^2 X} k_{\text{n}}$ & Necrotic lipid uptake rate & $1.4$ \\
    $\Tilde{k}_H$ & $\frac{a_0 \sigma_M}{\beta^2 X} k_{H}$ & Lipid efflux rate & $16$ \\
    $\Tilde{\delta}_S$ & $\frac{1}{\beta} \delta_S$  & Mediator natural decay rate & $1600^*$ \\
    $\Tilde{\alpha}$ & $\frac{a_0 \sigma_M}{S^{\text{c50}}\beta^2 X} \alpha$  & Resident mediator production per rLDL lipid & $850^*$ \\
    $\ell_\text{max}$ & $\ell_\text{max}$ & MDM maximum lipid capacity & $100$ \\
    \\
    \hdashline \\
    $\Tilde{\theta}$ & $\theta$ & LDL retention profile shape & $10$ \\
    $\Tilde{D}_{M}$ & $\frac{1}{\beta X^2} D_M$ & Diffusivity of unladen MDMs & $0.12$ \\
    $D_{\text{ap}}$ & $\frac{1}{\beta X^2} D_{\text{ap}}$  & Diffusivity of apoptotic cells & $0.03$  \\
    $D_{\text{n}}$ & $\frac{1}{\beta X^2} D_{\text{n}}$   & Diffusivity of necrotic cells & $0.03$ \\
    $\Tilde{D}_{\text{\scaleto{LDL}{3.5pt}}}$ & $\frac{1}{\beta X^2} D_{\text{\scaleto{LDL}{3.5pt}}}$ & Diffusivity of free LDL & $58$ \\
    $\Tilde{D}_{H}$ & $\frac{1}{\beta X^2} D_H$ & Diffusivity of HDL & $140$ \\
    $\Tilde{D}_{S}$ & $\frac{1}{\beta X^2} D_S$ & Diffusivity of mediators & $3600^*$ \\
    $\Tilde{\psi}_{D}$ & $\psi_D$ & Sensitivity of MDM diffusivity to lipid load & $\in[0,1]$ \\    
    $\Tilde{\psi}_{\beta}$ & $\psi_D$ & Sensitivity of MDM lifespan to lipid load & $\in[0,1]$ \\    
    \bottomrule
    \end{tabular} \label{tab: nondim parameters}
\end{table}
We define:
\begin{align} \label{eqn: nondim t,x}
     &\Tilde{x} := \frac{x}{X}, & &\Tilde{t} := \beta t,
\end{align}
so that spatial depth, $0 \leq \Tilde{x} \leq 1$, is measured relative to the width of the tunica intima, $X \approx$ 0.40 mm, and time, $\Tilde{t} \geq 0$, is measured in units of the mean lifespan of lipid-unladen MDMs, $\beta^{-1} \approx 1$ month. We scale the remaining variables as follows:
\begin{align}
    \begin{split} \label{eqn: nondim}
    &\Tilde{m}_{\ell}(\Tilde{x}, \Tilde{t}) := \frac{\beta X}{\sigma_M} m_{\ell}(x,t), \qquad \qquad \quad \, \, \, \Tilde{M}(\Tilde{x}, \Tilde{t}) := \frac{\beta X}{\sigma_M} M(x,t), \\
    &\Tilde{L}_M(\Tilde{x}, \Tilde{t}) := \frac{\beta X}{a_0 \sigma_M} L_M(x,t), \qquad \qquad \Tilde{\bm{L}}(\Tilde{x}, \Tilde{t}) := \frac{\beta X}{a_0 \sigma_M} \bm{L}(x, t), \\
    &\Tilde{H}(\Tilde{x}, \Tilde{t}) := \frac{\beta X}{a_0 \sigma_M} H(x, t), \qquad \qquad  \quad \, \,   \Tilde{S}(\Tilde{x}, \Tilde{t}) := \frac{1}{S^\text{c50}} S(x, t).
    \end{split}
\end{align}
Eqs. \eqref{eqn: nondim} scale the MDM densities relative to the maximum influx per (lipid-unladen) MDM lifespan, $\sigma_M X^{-1} \beta^{-1} \approx 17000$ mm$^{-3}$. The lipid densities and HDL capacity are measured relative to the maximum influx of MDM endogenous lipid per MDM lifespan, $a_0 \sigma_M X^{-1} \beta^{-1} \approx 45$  mg/dL. The inflammatory mediator density is scaled relative to the density for half-maximal MDM recruitment, $S^{\text{c50}} \approx 5$ ng/mL. We also introduce a number of dimensionless parameters in Table \ref{tab: nondim parameters}.

Applying the non-dimensionalisation \eqref{eqn: nondim} and definitions of Table \ref{tab: nondim parameters}, and dropping the tildes for notational convenience, we obtain the following dimensionless PDEs for the MDM population:
\begin{align}
    \begin{split} \label{eqn: ml nondim}
        \frac{\partial m_\ell}{\partial t} &= D_M \bigg[ 1 - \psi_D \Big( \frac{\ell}{\ell_\text{max}} \Big) \bigg] \frac{\partial^2 m_\ell}{\partial x^2} - \frac{ m_\ell}{1 - \psi_\beta \big( \frac{\ell}{\ell_\text{max}} \big)} \\
        &\quad + \bm{k_L} \cdot \bm{L} \, \, \big[ (\ell_\text{max}-\ell + 1) m_{\ell-1} - (\ell_\text{max} - \ell)m_{\ell} \big] \\
    &\quad + k_H \, \, \, \, H \big[ (\ell + 1) m_{\ell + 1} - \ell m_{\ell} \big], \\
    \end{split}
\end{align}
where:
\begin{align}
    m_{-1} = m_{\ell_\text{max}+1} = 0,
\end{align}
for every $\ell = 0, 1, \dots, \ell_\text{max}$. The remaining variables satisfy the PDEs:
\begin{align}
    \frac{\partial M}{\partial t} &= D_M \frac{\partial^2}{\partial x^2} \bigg[ M - \psi_D \sum_{\ell = 0}^{\ell_\text{max}} \Tilde{\ell} m_\ell \bigg] - \bigg[ M + \psi_\beta \sum_{\ell = 0}^{\ell_\text{max}} \frac{\Tilde{\ell} m_\ell}{1 - \psi_\beta \Tilde{\ell} } \bigg], \label{eqn: M nondim} \\
    \begin{split}
        \frac{\partial L_M}{\partial t} &= D_M \frac{\partial^2}{\partial x^2} \bigg[ L_M - \psi_D \sum_{\ell = 0}^{\ell_\text{max}}  ( 1 + \kappa \Tilde{\ell}) \Tilde{\ell} m_\ell  \bigg]  + \bm{k_L} \cdot \bm{L} [(\kappa + 1) M - L_M] \\ &\qquad \quad - \, \, \, \, \, \bigg[ L_M +  \psi_\beta \sum_{\ell = 0}^{\ell_\text{max}} \frac{(1 + \kappa \Tilde{\ell}) \Tilde{\ell} m_\ell}{1  - \psi_\beta \Tilde{\ell} } \bigg]  - k_H \,  H (L_M - M),
    \end{split} \label{eqn: L_M nondim} \\
    \begin{split}
        \frac{\partial L_{\text{\scaleto{LDL}{3.5pt}}}}{\partial t} &= D_{\text{\scaleto{LDL}{3.5pt}}} \frac{\partial^2 L_{\text{\scaleto{LDL}{3.5pt}}}}{\partial x^2} - k_b L_{\text{\scaleto{LDL}{3.5pt}}} [K_\text{r}(x) - L_\text{r}] + k_{-b}L_r \\
        &\quad - k_{\text{\scaleto{LDL}{3.5pt}}} L_{\text{\scaleto{LDL}{3.5pt}}} [(\kappa + 1) M - L_M],
    \end{split} \label{eqn: lldl nondim} \\
    \begin{split}
        \frac{\partial L_\text{r}}{\partial t} &= k_b L_{\text{\scaleto{LDL}{3.5pt}}} [K_\text{r}(x) - L_\text{r}] - k_{-b}L_r  - k_\text{r} L_\text{r} [(\kappa + 1) M - L_M],
    \end{split} \\
    \begin{split}
        \frac{\partial L_\text{ap}}{\partial t} &= D_\text{ap} \frac{\partial^2 L_\text{ap}}{\partial x^2} + \bigg[ L_M +  \psi_\beta \sum_{\ell = 0}^{\ell_\text{max}} \frac{(1 + \kappa \Tilde{\ell} ) \Tilde{\ell} m_\ell}{1  - \psi_\beta \Tilde{\ell} } \bigg] \\
        &\quad - \nu L_\text{ap} - k_\text{ap}L_\text{ap} [(\kappa + 1) M - L_M],
    \end{split} \\
    \frac{\partial L_\text{n}}{\partial t} &= D_\text{n} \frac{\partial^2 L_\text{n}}{\partial x^2} +  \nu L_\text{ap} - k_\text{n} L_\text{n} [(\kappa + 1) M - L_M], \\
    \frac{\partial H}{\partial t} &= D_H \frac{\partial^2 H}{\partial x^2} - k_H H (L_M - M), \label{eqn: H nondim}\\
    \frac{\partial S}{\partial t} &= D_S \frac{\partial^2 S}{\partial x^2} + \alpha L_\text{r} - \delta_S S, \label{eqn: S nondim}
\end{align}
where $\Tilde{\ell} := \ell/\ell_\text{max}$ and 
\begin{align} \label{eqn: Kr nondim}
    K_\text{r}(x):= \bar{K}_\text{r} \cdot \bigg( \frac{2}{1 + e^{\theta (1/2-x)}} \bigg).
\end{align} We impose the following boundary conditions:
\begin{align} 
    -D_M \frac{\partial m_\ell}{\partial x} &= 
    \begin{cases} \label{eqn: ml bconds nondim}
         \frac{S}{1 + S} \cdot \delta_{\ell, 0} - \gamma m_{\ell} & \text{at } x=0; \\
         \gamma m_{\ell} & \text{at } $x=1$,
    \end{cases} \qquad \forall \ell = 0, 1, \dots \ell_\text{max},  \\
    -D_M \frac{\partial M}{\partial x} &= 
    \begin{cases}
             \frac{S}{1 + S} - \gamma M & \text{at } x=0; \\
         \gamma M & \text{at } $x=1$,
    \end{cases} \label{eqn: M bconds nondim}\\
        -D_M \frac{\partial L_M}{\partial x} &= 
    \begin{cases}
             \frac{S}{1 + S} - \gamma L_M & \text{at } x=0; \\
         \gamma L_M & \text{at } $x=1$,
    \end{cases} \label{eqn: L_M bconds nondim} \\
        -D_{\text{\scaleto{LDL}{3.5pt}}} \frac{\partial L_{\text{\scaleto{LDL}{3.5pt}}}}{\partial x} &= 
    \begin{cases}
        P_L^{(0)} (L^\star - L_{\text{\scaleto{LDL}{3.5pt}}}) & \text{at } x = 0; \\
        P_L^{(1)} (L_{\text{\scaleto{LDL}{3.5pt}}} - 0) & \text{at } x = 1,
    \end{cases} \label{eqn: LDLbconds nondim} \\
    \frac{\partial L_\text{ap}}{\partial x} &= \frac{\partial L_\text{n}}{\partial x} = 0 \quad  \text{at } x = 0, 1, \label{eqn: Lap Ln bconds nondim} \\
        -D_H \frac{\partial H}{\partial x} &= 
    \begin{cases}
        P_H^{(0)} (H^\star - H) & \text{at } x = 0; \\
        P_H^{(1)} (H - 0) & \text{at } x = 1,
    \end{cases} \label{eqn: Hbconds nondim} \\
            -D_S \frac{\partial S}{\partial x} &=    
            \begin{cases}
        P_S^{(0)} (0 - S) & \text{at } x = 0; \\
        P_S^{(1)} (S - 0) & \text{at } x = 1,
    \end{cases} \label{eqn: Sbconds nondim}
\end{align}
and close the model with the following initial conditions at $t=0$:
\begin{align}
    \begin{split}
        &m_\ell = 0 \quad \forall \ell = 0, 1, \dots, \ell_\text{max}, \\
        &M = L_M = L_{\text{\scaleto{LDL}{3.5pt}}} = L_\text{r} = L_\text{ap} = L_\text{n} = H = S = 0. 
    \end{split} \label{eqn: init nondim}
\end{align}

\subsection{Separation of timescales} \label{sec: separation of timescales}

The parameters in Table \ref{tab: nondim parameters} span several orders of magnitude. In particular, the mediator parameters $\delta_S$, $\alpha$ and $D_S$ are substantially larger than the other constants. This discrepancy means that numerical simulations require small timesteps to maintain stability. To improve numerical efficiency and derive analytical insight into model behaviour, we approximate the mediator dynamics via separation of timescales. Accordingly, we define:
\begin{align}
    &\epsilon := \delta_S^{-1} = 6.25 \times 10^{-4} \ll 1, & &\alpha = \epsilon^{-1} \hat{\alpha}, & &D_S = \epsilon^{-1} \hat{D}_S,
\end{align}
and consider the limit $\epsilon \rightarrow 0$ where $\hat{\alpha}$, $\hat{D}_S = \mathcal{O}(1)$ as $\epsilon \rightarrow 0$. In this limit, Eqs. \eqref{eqn: S nondim} and \eqref{eqn: Sbconds nondim} become:
\begin{align}
    \epsilon \frac{\partial S}{\partial t} &= \hat{D}_S \frac{\partial^2 S}{\partial x^2} + \hat{\alpha} L_r - \hat{\delta}_S S, &  -\hat{D}_S \frac{\partial S}{\partial x} &=    
            \epsilon \begin{cases}
        P_S^{(0)} (0 - S) & \text{at } x = 0; \\
        P_S^{(1)} (S - 0) & \text{at } x = 1.
    \end{cases} \label{eqn: S asymp}
\end{align}
Substituting the expansion $S = S^{(0)} + \epsilon S^{(1)} + \dots $ into Eqs. \eqref{eqn: S asymp} and balancing the leading order terms gives:
\begin{align} \label{eqn: S0}
    0 &= \hat{D}_S \frac{\partial^2 S}{\partial x^2}^{\!(0)} + \hat{\alpha} L_r - \hat{\delta}_S S^{(0)}, & \frac{\partial S}{\partial x}^{\! (0)} &= 0 \quad \text{at } x = 0,1.
\end{align}
Hence, $S$ satisfies a quasi-steady approximation with no-flux boundary conditions at leading order. Eqs. \eqref{eqn: S0} can be solved in terms of $L_\text{r}(x,t)$ using variation of parameters: 
\begin{align}
    \begin{split} \label{eqn: S(x,t)}
        S^{(0)}(x,t) &= \frac{\hat{\alpha}}{\sqrt{\hat{D}_S}} \bigg[ \int_{x}^{1} L_r(\xi, t)  \sinh \bigg( \frac{x - \xi}{\sqrt{\hat{D}_S}} \bigg) d\xi \\
        &\quad + \text{csch} \big(\frac{1}{\sqrt{\hat{D}_S}}\big) \cosh \bigg( \frac{1-x}{\sqrt{\hat{D}_S}} \bigg)  \int_{0}^{1} L_\text{r}(\xi, t)  \cosh \bigg( \frac{\xi}{\sqrt{\hat{D}_S}} \bigg) d\xi \bigg].
    \end{split}
\end{align}
Since $S$ is only coupled to the model via its value at $x = 0$ in Eqs. \eqref{eqn: ml bconds nondim}-\eqref{eqn: L_M bconds nondim}, we substitute $x = 0$ into Eq. \eqref{eqn: S(x,t)} and simplify to obtain:
\begin{align}
    S(0,t) \sim\frac{\hat{\alpha}}{\sqrt{\hat{D}_S} \sinh \Big(\frac{1}{\sqrt{\hat{D}_S}} \Big)} \int_0^1 L_\text{r}(x,t)\cdot \cosh \bigg( \frac{1-x}{\sqrt{\hat{D}_S}} \bigg) dx, \quad \text{as } \epsilon \rightarrow 0. \label{eqn: S(0,t)}
\end{align}
Eq. \eqref{eqn: S(0,t)} shows that, at leading order, the mediator density at the endothelium is a weighted integral of the rLDL density across the lesion. The rLDL density closer to $x = 0$ is weighted more strongly than that near $x=1$ to account for the decay of mediators as they diffuse towards the endothelium. \\



\subsection{Numerical scheme}

We use the method of lines, with spatial discretisation $\Delta x = 0.02$, to obtain numerical solutions of the reduced model, which comprises PDEs \eqref{eqn: ml nondim}-\eqref{eqn: H nondim},  boundary conditions \eqref{eqn: M bconds nondim}-\eqref{eqn: Hbconds nondim}, initial conditions \eqref{eqn: init nondim} and expression \eqref{eqn: S(0,t)} for $S(0,t)$. We approximate the spatial derivatives with a second-order central finite difference, and the integral in Eq. \eqref{eqn: S(0,t)} with the Trapezoidal Rule. We solve the resulting ODE system with the \textit{NDSolve} routine in Wolfram Mathematica. {\color{black}
We tested the validity of numerical solutions by verifying that the output of \textit{NDSolve} was robust to decreases in $\Delta x$ (i.e. the spatial discretisation), `MaxStepSize' (i.e. the temporal discretisation), increases in `PrecisionGoal' and `AccuracyGoal', and consistent with the steady state solutions for $\psi_\beta = \psi_D = 0$ in Sect. \ref{sec: lipid-independent}.
}

\section{Results} \label{sec: results}

We present the results in four sections.  We consider the case $\psi_D = \psi_\beta = 0$ in Sect. \ref{sec: lipid-independent}. Our analysis shows that this model, with lipid-independent MDM mobility and apoptosis, captures the overall time course of lesion development, 
but does not qualitatively reproduce the spatial densities seen in the Nakashima images (Fig. \ref{fig: Nakashima}). We then examine the impact at steady state of lipid-dependent apoptosis alone, $\psi_\beta > 0 = \psi_D$, and  lipid-dependent mobility alone, $\psi_D > 0 = \psi_\beta$,  in Sects. \ref{sec: lipid_dependent_apoptosis} and \ref{sec: lipid_dependent_diffusion} respectively. Our results show that lipid-dependent apoptosis alone cannot reproduce the observed spatial densities, but high sensitivities to  lipid-dependent mobility are sufficient. Finally, in Sect. \ref{sec: lipid_dependent_both} we consider the model output when both apoptosis and diffusion are lipid-dependent: $\psi_\beta, \psi_D > 0$.

\subsection{Lipid-independent MDM kinetics: $\psi_D = \psi_\beta = 0$} \label{sec: lipid-independent}

Results from a typical simulation for the case $\psi_D = \psi_\beta = 0$ are presented in Fig. \ref{fig: dynamics_ConstantRates}. We fix $L^\star = 4$ and $H^\star = 0.5$ so that LDL lipid dominates HDL capacity in the blood. The plots in column (a) show the components of the total lipid density:
\begin{align}
    L_\text{tot}(x,t) := L_{\text{\scaleto{LDL}{3.5pt}}}(x,t) + L_\text{r}(x,t) + L_M(x,t) + L_\text{ap}(x,t) + L_\text{n}(x,t), \label{eqn: Ltot}
\end{align}
while the quantities in plot (e) are the mean lipid density and HDL lipid capacity, respectively given by
\begin{align} \label{eqn: Ltot H mean}
    &\overline{L}_\text{tot}(t) := \int_0^1 L_\text{tot}(x,t) dx, & &\overline{H}_\text{tot}(t) := \int_0^1 H(x,t) dx.
\end{align}

\begin{figure}
    \centering
    \includegraphics[width=0.99\textwidth]{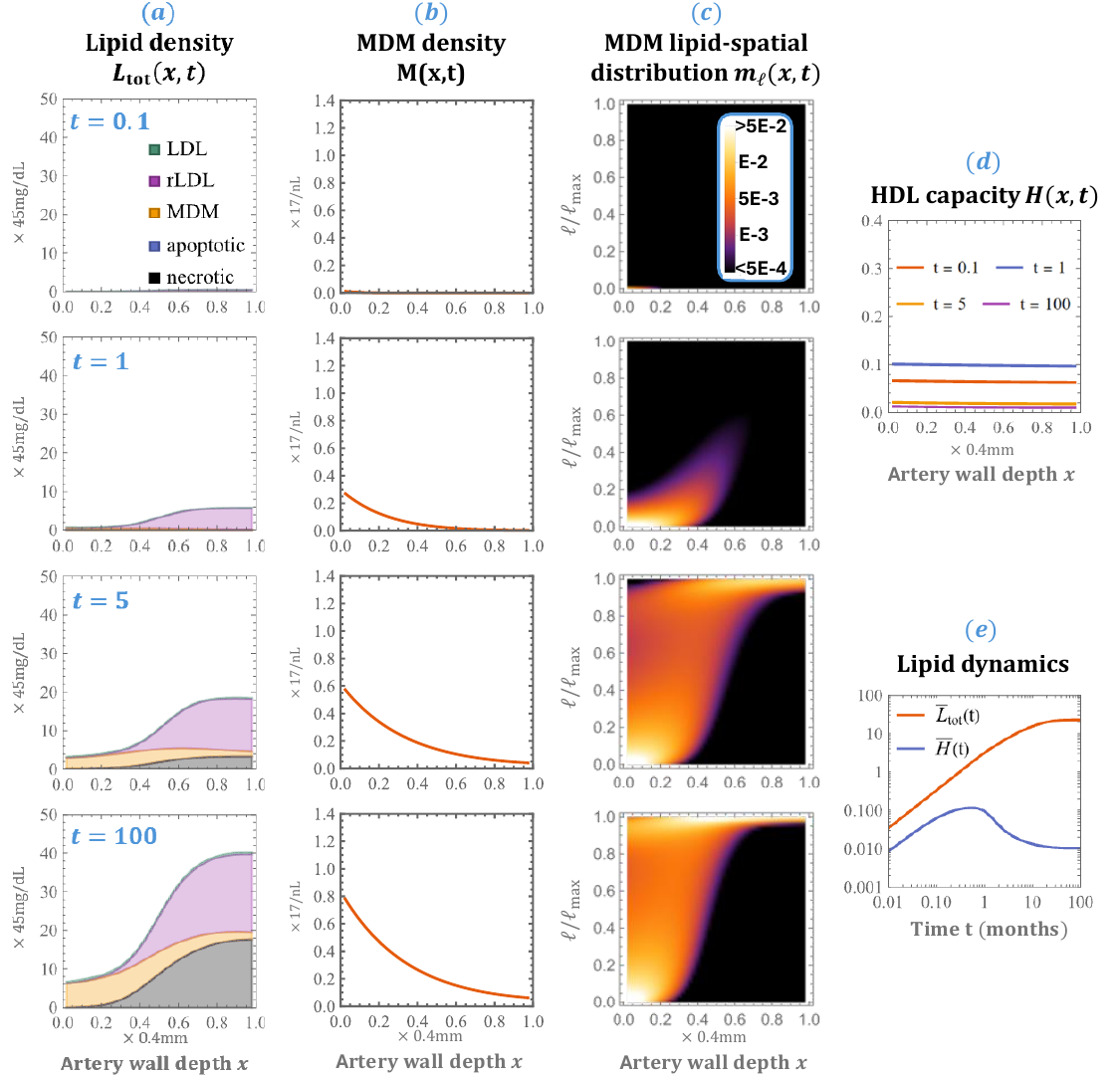}
    \caption{\textbf{Time evolution of the model lesion  when MDM mobility and apoptosis are independent of lipid content: $\mathbf{\psi_D = \psi_\beta = 0}$.} Columns (a), (b) and (c) respectively show the lipid densities, MDM density and MDM lipid-spatial distribution at times $t = 0.1$, $1$, $5$ and $100$. Plot (d) shows the spatial uniformity of HDL capacity. Finally, plot (e) shows the time-evolution of the mean total lipid density and HDL capacity, as defined by Eqs. \eqref{eqn: Ltot H mean}. Note how, in contrast to the Nakashima images, the lipid density monotonically increases with $x$, and the MDM density monotonically decreases with $x$ for all $t > 0$. We use $L^\star = 4$ and $H^\star = 0.5$.}
    \label{fig: dynamics_ConstantRates}
\end{figure}


The time evolution of the lipid densities is shown in Fig. \ref{fig: dynamics_ConstantRates}(a). At early times ($t \leq 1$), the lesion accumulates rLDL deep in the wall. This behaviour, consistent with the Nakashima images (Fig. \ref{fig: Nakashima}), occurs due to the influx of free LDL at $x = 0$ and its subsequent binding to ECM proteoglycans. The high diffusivity of free LDL ($D_\text{LDL} = 57$), ensures that spatial variation in free LDL is small and so the spatial profile of rLDL is dominated by the non-uniform retention capacity, $K_\text{r}(x)$, defined in Eq. \eqref{eqn: Kr nondim}. Moreover, the proportion of free LDL to rLDL is very small due to the large mean retention capacity ($\overline{K}_\text{r} = 25$), relative to blood LDL density ($L^\star = 4$). In the months following LDL accumulation ($1 \leq t \leq 5$), the MDM lipid density grows markedly due to rLDL uptake. The lesion also accumulates apoptotic and necrotic lipid at this stage due to MDM death. The ratio of apoptotic to necrotic lipid is also small, due mainly to the high secondary necrosis rate, $\nu = 37$. Finally, the model tends to a nonzero steady state as $t \rightarrow \infty$. The lesion is approximately at steady state within $t=100$ MDM lifespans (see Fig. \ref{fig: dynamics_ConstantRates}(e), noting the logarithmic scale). The lesion continues to accumulate necrotic lipid deep in the intima during this final stage. 

The time evolution of the MDM density is given in Fig. \ref{fig: dynamics_ConstantRates}(b). The MDM density grows over time and maintains a monotonically decreasing spatial profile. As expected, $M(x,t)$ attains its maximum value at $x = 0$, where MDMs enter the lesion, and fixing $\psi_D = \psi_\beta = 0$ ensures that their movement and death are decoupled from lipid content. We note that $M(x,t)$ decreases with $x$ sufficiently smoothly that $M(1,t) > 0$ for $t$ large; indeed $M(1,100) \approx 0.1 \times M(0,100)$ for the simulation in Fig. \ref{fig: dynamics_ConstantRates}. Hence, by contrast to the Nakashima images, the model predicts that a substantial portion of MDMs eventually emigrate to the tunica media. 

The dynamics of the lipid-spatial MDM distribution, $m_\ell(x,t)$, are shown in Fig. \ref{fig: dynamics_ConstantRates}(c). At early times, $t \leq 1$, the distribution grows in a wave-like manner from the origin towards higher values of $\ell$ and $x$. This behaviour reflects the early increase in MDM lipid content due to rLDL uptake and concurrent diffusion into the lesion. As $t$ increases, the distribution stretches towards $(x, \ell/ \ell_\text{max}) = (1,1)$, reflecting a growth in the number of lipid-laden MDMs deep in the lesion. We note that $m_\ell(x,t)$ is concentrated on higher values of $\ell$ as $x$ increases; MDMs deeper in the lesion typically contain a higher lipid content than those nearer the endothelium. For $5 \leq t \leq 100$ the MDM distribution also extends towards $(x, \ell/ \ell_\text{max}) = (0,1)$. This behaviour is partially due to the diffusion of lipid-laden MDMs from deeper in the lesion back towards the endothelium as the model tends to steady state. 

Fig. \ref{fig: dynamics_ConstantRates}(d) shows that the HDL capacity, $H(x,t)$, is approximately spatially uniform for all $t > 0$. This is due to the large HDL diffusivity, $D_H = 140$. Indeed, if $D_H = \epsilon^{-1} \hat{D}_H$ for some $\hat{D}_H = \mathcal{O}(1)$ as $\epsilon \rightarrow 0$ (recall $\epsilon := \delta_S^{-1} \ll 1$ ), then Eqs. \eqref{eqn: H nondim} and \eqref{eqn: Hbconds nondim} become
\begin{align}
    \epsilon \frac{\partial H}{\partial t} &= \hat{D}_H \frac{\partial^2 H}{\partial x^2} - \epsilon k_H H (L_M - M) \label{eqn: H asymp} \\
    - \hat{D}_H \frac{\partial H}{\partial x} &= \epsilon \begin{cases} 
        P_H^{(0)} (H^\star - H) & \text{at } x = 0; \\
        P_H^{(1)} H & \text{at } x = 1.
    \end{cases} \label{eqn: H asymp bconds}
\end{align}
Substituting $H = H^{(0)} + \epsilon H^{(1)} + \dots $ into Eqs. \eqref{eqn: H asymp} and \eqref{eqn: H asymp bconds}, and balancing the leading order terms shows that 
\begin{align} \label{eqn: H spatial uniform}
    &\frac{\partial^2 H^{(0)}}{\partial x^2} = 0, \quad \frac{\partial H^{(0)}}{\partial x} \bigg\vert_{x=0,1} = 0, & &\implies  H^{(0)} = H^{(0)}(t).
\end{align}
Hence, the HDL lipid capacity is spatially uniform at leading order. To derive the dynamics of $H^{(0)}(t)$, we balance the $\mathcal{O}(\epsilon)$ terms in Eq. \eqref{eqn: H asymp} and integrate with respect to $x \in (0,1)$. Doing so gives
\begin{align}
    \int_{0}^{1} \frac{\partial H^{(0)}}{\partial t} dx &= \hat{D}_{H} \frac{\partial H^{(1)}}{\partial x} \bigg\vert_{x=0}^{x=1} - k_H H^{(0)} \int_{0}^{1} (L_M - M) dx. 
\end{align}
Using result Eq. \eqref{eqn: H spatial uniform} to simplify the left-hand side, and Eq. \eqref{eqn: H asymp bconds} at $\mathcal{O}(\epsilon)$ to evaluate the boundary terms on the right-hand side, we find that
\begin{align} \label{eqn: H asymptotic1}
    H(x,t) \sim H^{(0)}(t) \quad \text{as } \epsilon \rightarrow 0,
\end{align}
where
\begin{align} \label{eqn: H asymptotic2}
    \begin{split}
        \frac{dH^{(0)}}{dt} &= P_H^{(0)} (H^\star -H^{(0)}) - P_H^{(1)} H^{(0)} - k_H H^{(0)} \int_0^1 (L_M - M)dx.
    \end{split}
\end{align}
Eqs. \eqref{eqn: H asymptotic1} and \eqref{eqn: H asymptotic2} express that, at leading order, the HDL density is spatially uniform and evolves according to the rates of exchange with the bloodstream, tunica media and total efflux by MDMs across the lesion $x \in (0,1)$. These mechanisms give rise to the non-monotonic dynamics seen in Fig. \ref{fig: dynamics_ConstantRates}(e); HDL capacity initially grows from zero due to influx from the blood, and decreases once MDMs have ingested enough lipid that the combination of efflux and loss to the tunica media exceeds influx from the bloodstream. By contrast, the mean lipid density across the lesion increases monotonically with $t$; the lesion becomes more lipid laden with time. \\

\noindent \textbf{Steady state. } Typical steady state solutions when $\psi_D = \psi_\beta = 0$ are shown in Fig. \ref{fig: steady1_ConstantRates}. Again, we note that the steady state MDM density decreases monotonically with $x$. We prove this result by direct calculation. At steady state, Eqs. \eqref{eqn: M nondim} and \eqref{eqn: M bconds nondim} become
\begin{align} \label{eqn: M steady bvp}
    &0 = D_M \frac{d^2 M}{d x^2} - M, & &-D_M \frac{d M}{d x} = 
    \begin{cases}
             \frac{S_\infty}{1 + S_\infty} - \gamma M & \text{at } x=0; \\
         \gamma M & \text{at }  x=1,
    \end{cases}
\end{align}
where $S_\infty := \lim_{t \rightarrow \infty} S(0,t)$ is the limiting value of Eq. \eqref{eqn: S(0,t)}. The boundary value problem \eqref{eqn: M steady bvp} admits the exact solution
\begin{align} \label{eqn: M_inf(x)}
    M (x) = A \cosh \Bigg( \frac{1-x}{\sqrt{D_M}} \Bigg) + B \sinh \Bigg( \frac{1-x}{\sqrt{D_M}} \Bigg),
\end{align}
where $A$ and $B$ are positive constants:
\begin{align}
    A &= \frac{S_\infty}{1 + S_\infty} \cdot \frac{\sqrt{D_M}}{2 \gamma \sqrt{D_M} \cosh(1/\sqrt{D_M}) + (D_M + \gamma^2) \sinh(1/\sqrt{D_M})}, \\
    B &= \frac{\gamma A}{\sqrt{D_M}}.
\end{align}
Differentiating Eq. \eqref{eqn: M_inf(x)} then gives
\begin{align}
    M'(x) = - \frac{A}{\sqrt{D_M}} \sinh \Bigg( \frac{1-x}{\sqrt{D_M}} \Bigg) - \frac{B}{\sqrt{D_M}} \cosh \Bigg( \frac{1-x}{\sqrt{D_M}} \Bigg) < 0,
\end{align}
where the inequality follows from $A,B > 0$ and $1-x \geq 0$. Hence, $M(x)$ is monotone decreasing and, consequently, our model with lipid-independent MDM kinetics $\psi_D = \psi_\beta = 0$ cannot qualitatively replicate the MDM densities found in the Nakashima images. 

\begin{figure}
    \centering
    \includegraphics[width=0.99\textwidth]{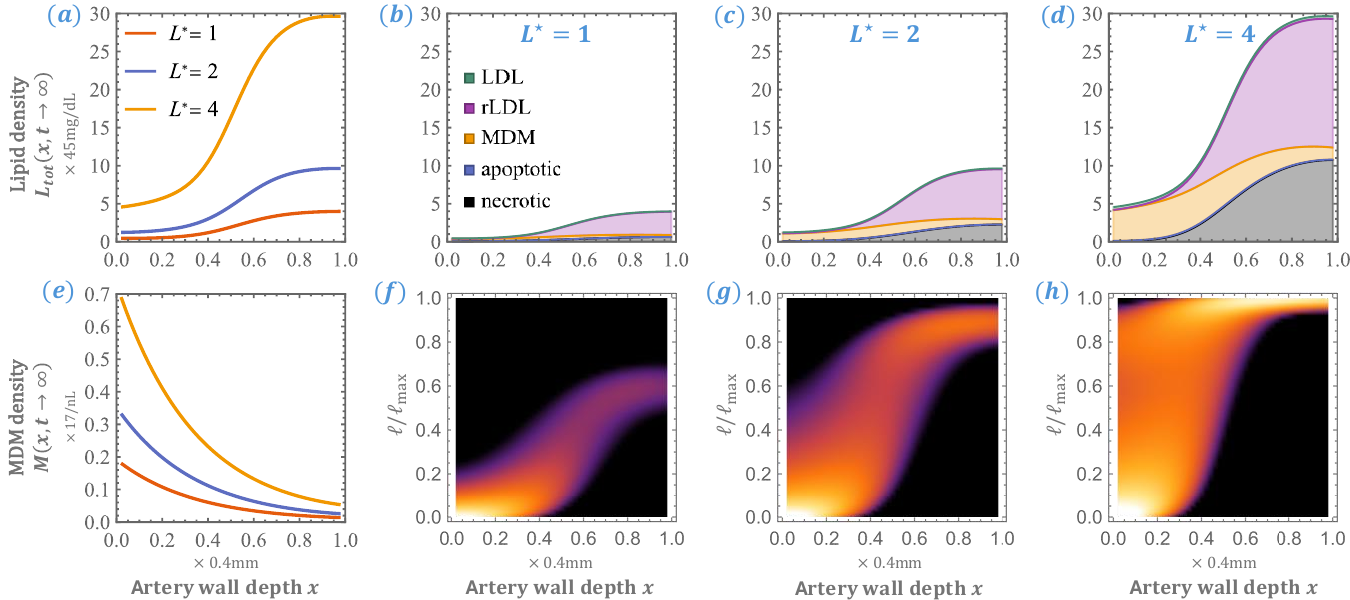}
    \caption{\textbf{Steady state lesion composition when MDM mobility and apoptosis are independent of lipid content: $\psi_D = \psi_\beta = 0$.} The plots depict solutions to the model equations at steady state for the blood LDL densities $L^\star = 1$, $2$ and $4$, and HDL capacity $H^\star = 1$. The top row depicts the lipid densities; plot (a) is the total lipid density, and plots (b)-(d) show the densities of the lipid components. The bottom row shows the MDM densities; plot (e) is the total MDM density, and plots (f)-(h) show the lipid-spatial MDM distribution, using the same colour scale as in Fig. \ref{fig: dynamics_ConstantRates}. Note how the MDM density decreases monotonically with $x$, in contrast to the Nakashima images.}
    \label{fig: steady1_ConstantRates}
\end{figure}

\begin{figure}
    \centering
    \includegraphics[width=0.99\textwidth]{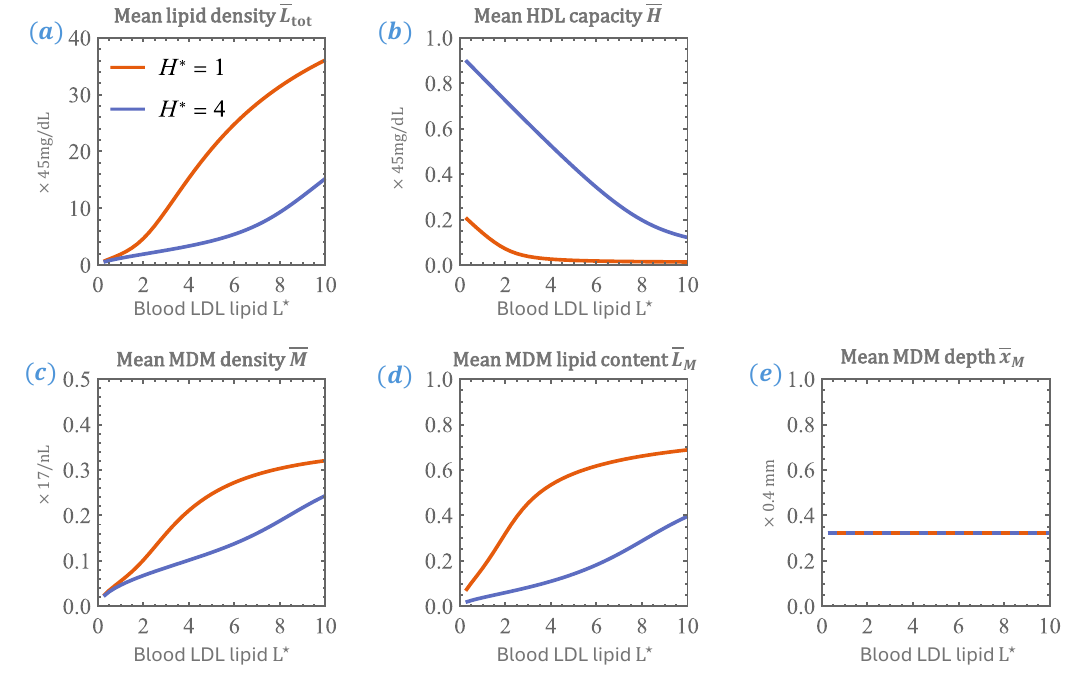}
    \caption{\textbf{Steady state summary statistics when MDM mobility and apoptosis are independent of lipid content:  $\psi_D = \psi_\beta = 0$.} Plots (a) and (b) show the mean total lipid density and HDL capacity, defined in Eqs. \eqref{eqn: Ltot H mean}. Plots (c)-(e) show the mean MDM density, (normalised) lipid content and depth, which are defined in Eqs. \eqref{eqn: MDM means}. Note how mean MDM depth is independent of both $L^\star$ and $H^\star$, by contrast to the other quantities. }
    \label{fig: steady2_ConstantRates}
\end{figure}

Since we use the results of our model with $\psi_D = \psi_\beta = 0$ for comparison in future sections, we continue to explore its steady state properties. Fig. \ref{fig: steady2_ConstantRates} shows how various summary statistics depend on blood LDL content.
Plots (a) and (b) respectively show the mean total lipid density, $\overline{L}_\text{tot}$ and HDL capacity, $\overline{H}$, as defined by Eqs \eqref{eqn: Ltot H mean}. As expected, $\overline{L}_\text{tot}$ increases with $L^\star$. By contrast, $\overline{H}$ decreases with $L^\star$ due to increased rates of MDM efflux. Plots (c), (d) and (e) respectively show the mean MDM density, $\overline{M}$, lipid content, $\overline{L}_M$, and depth, $\overline{x}_M$, which are defined by
\begin{align} \label{eqn: MDM means}
    &\overline{M} := \int_0^1 M dx, & &\overline{L}_M := \frac{1}{\kappa \overline{M}} \int_0^1 (L_M - M) dx, & &\overline{x}_M := \frac{1}{\overline{M}} \int_0^1 xM dx.
\end{align}
We note that $0 \leq \overline{L}_M \leq 1$ is normalised such that $\overline{L}_M = 0$ when all MDMs have no ingested lipid and $\overline{L}_M = 1$ when all MDMs are at maximal capacity. Both $\overline{M}$ and $\overline{L}_M$ increase with $L^\star$, whereas $\overline{x}_M$ appears to be independent of $L^\star$ and $H^\star$. Indeed, substituting the exact solution \eqref{eqn: M_inf(x)} into Eq. \eqref{eqn: MDM means} shows that
\begin{align} \label{eqn: xM}
    \overline{x}_M = \frac{D_M^2 + \gamma - D_M \big( D_M \cosh(1/D_M)+\gamma \sinh(1/D_M) \big)}{\gamma - \gamma \cosh(1/D_M) - D_M \sinh(1/D_M)}.
\end{align}
Hence, when $\psi_D = \psi_\beta = 0$, the mean MDM depth only depends on the diffusivity, $D_M$, and egress rate, $\gamma$.

\subsection{Lipid-dependent apoptosis only:  $\psi_D = 0$, $\psi_\beta > 0$} \label{sec: lipid_dependent_apoptosis}

We next investigate whether lipid-dependent MDM apoptosis alone is sufficient to qualitatively replicate the Nakashima images. Hence, we set $\psi_\beta \geq 0$ and continue to fix $\psi_D = 0$ so that MDM mobility is independent of lipid content. 

\begin{figure}[h]
    \centering
    \includegraphics[width=0.99\textwidth]{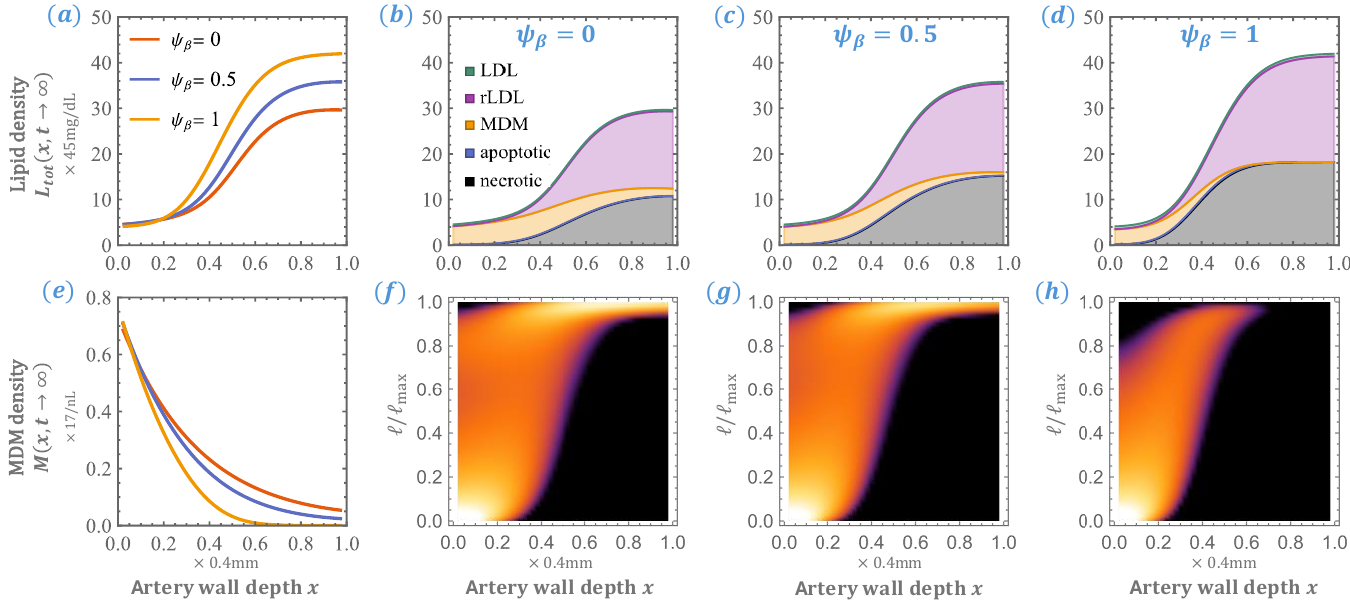}
    \caption{\textbf{Steady state lesion composition when MDM apoptosis increases with lipid content: $\mathbf{\psi_\beta > 0, \psi_D = 0}$.} We compute steady state solutions for $\psi_\beta = 0, 0.5, 1$, with $L^\star = 4$, $H^\star = 1$. Plot (a) shows the total lipid density and (b)-(d) the densities of the different lipid sub-types. Plot (e) depicts the density of MDMs and (f)-(h) the lipid-spatial MDM distribution, using the same colour scale as in Fig. \ref{fig: dynamics_ConstantRates}. Note how increases in $\psi_\beta$ yield a marked decline in MDM density deep in the lesion.}
    \label{fig: steady1_PsiB}
\end{figure}

Fig. \ref{fig: steady1_PsiB} shows typical steady state solutions for $\psi_\beta = 0$, $0.5$ and $1$. Plot (a) illustrates that increasing $\psi_\beta$ increases the total lipid density, particularly deep in the lesion. By comparing plots (c) and (d) to (b), we note that the increase in lipid content reflects a rise in necrotic and rLDL lipid towards $x = 1$, despite smaller quantities of MDM lipid. Plot (e) shows that the decline in MDM lipid is not due to low lipid loads, but rather a reduction in MDM density towards $x = 1$. Indeed, plots (f)-(h) indicate that MDM lipid content increases with depth for $\psi_\beta = 0, 0.5, 1$. Together, these observations suggest that lipid-dependent MDM apoptosis drives increased rates of MDM death deep in the lesion, where lipid loads are greatest, causing buildup of necrotic and rLDL lipid. Since lipid-laden MDMs are more likely to die than nascent MDMs, there is less MDM lipid in the lesion and, hence, a lower rate of lipid efflux to HDL. The decreased efflux rate means that more of the lipid which enters the lesion via LDL remains in the lesion, increasing the total lipid burden.

\begin{figure}[h]
    \centering
    \includegraphics[width=0.99\textwidth]{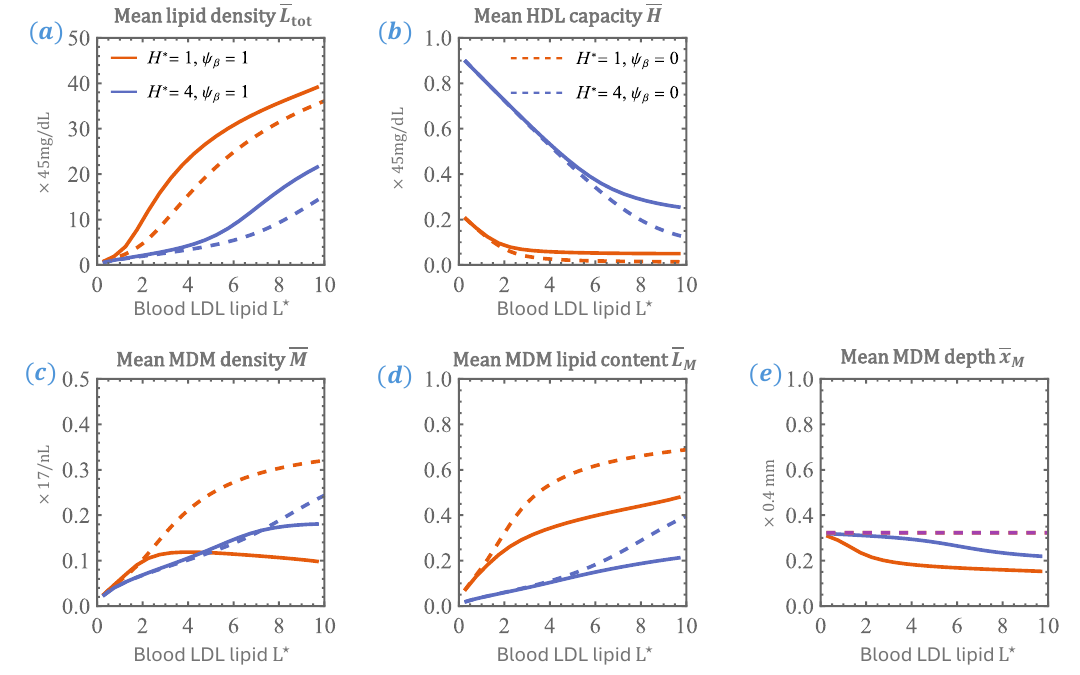}
    \caption{\textbf{Steady state summary statistics when MDM apoptosis increases with lipid content (solid) vs. lipid-independent apoptosis (dashed). } Plots (a) and (b) show how mean lipid density and HDL capacity, defined by Eqs. \eqref{eqn: Ltot H mean}, are higher when $\psi_\beta = 1$. Plots (c), (d) and (e) respectively depict the reduction in mean MDM density, lipid content, and depth for $\psi_\beta = 1$. Note how the differences between the outputs for $\psi_\beta = 1$ and $\psi_\beta = 0$ are greater for higher values of blood LDL content, $L^\star$, and lower values of blood HDL capacity, $H^\star$. }
    \label{fig: steady2_PsiB}
\end{figure}

We further highlight the impact of lipid-dependent apoptosis in Fig. \ref{fig: steady2_PsiB} where we compare summary statistics for $\psi_\beta = 1$ to the lipid-independent case $\psi_\beta = 0$. Consistent with Fig. \ref{fig: steady1_PsiB}, plot (a) shows that $\psi_\beta = 1$ yields a greater mean lipid density, $\overline{L}_\text{tot}$, than $\psi_\beta = 0$ for all $L^\star \in (0,10)$. Plot (b) shows that mean HDL capacity is also greater when $\psi_\beta = 1$, confirming a decline in HDL efflux when apoptosis depends on lipid content. Plots (c)-(e) show that mean MDM density, $\overline{M}$, lipid content, $\overline{L}_M$, and depth, $\overline{x}_M$, are all reduced relative to the case of lipid-independent death. The reduction is greater for high blood LDL densities, $L^\star$, where MDM lipid loads are greater and consequently rates of apoptosis. 

Importantly, as suggested by Fig. \ref{fig: steady1_PsiB}(e), we note that the MDM densities for $\psi_\beta > 0 = \psi_D$ cannot exhibit a local maximum at steady state. Indeed, rearranging Eq. \eqref{eqn: M nondim} at steady state gives
\begin{align} \label{eqn: M concavity}
    \frac{d^2 M}{dx^2} = \frac{1}{D_M} \bigg[ M + \psi_\beta \sum_{\ell = 0}^{\ell_\text{max}} \frac{\ell m_{\ell}}{ \ell_\text{max} - \psi_\beta \ell} \bigg] \geq 0,
\end{align}
where the inequality follows from $\psi_\beta < 1$ (the summand is non-negative). Hence, $M(x)$ is concave-up wherever it is nonzero and cannot attain an internal, local maximum. We conclude that introducing lipid-dependent apoptosis only to the model is not sufficient to qualitatively reproduce the Nakashima images.

\subsection{Lipid-dependent mobility only: $\psi_D > 0$, $\psi_\beta = 0$} \label{sec: lipid_dependent_diffusion}

Here we examine whether introducing lipid-dependent MDM mobility only is sufficient to qualitatively replicate the Nakashima images. Accordingly, we set $\psi_D > 0$ and assume $\psi_\beta = 0$ so that MDM apoptosis is independent of lipid content. 

\begin{figure}
    \centering
    \includegraphics[width=0.99\textwidth]{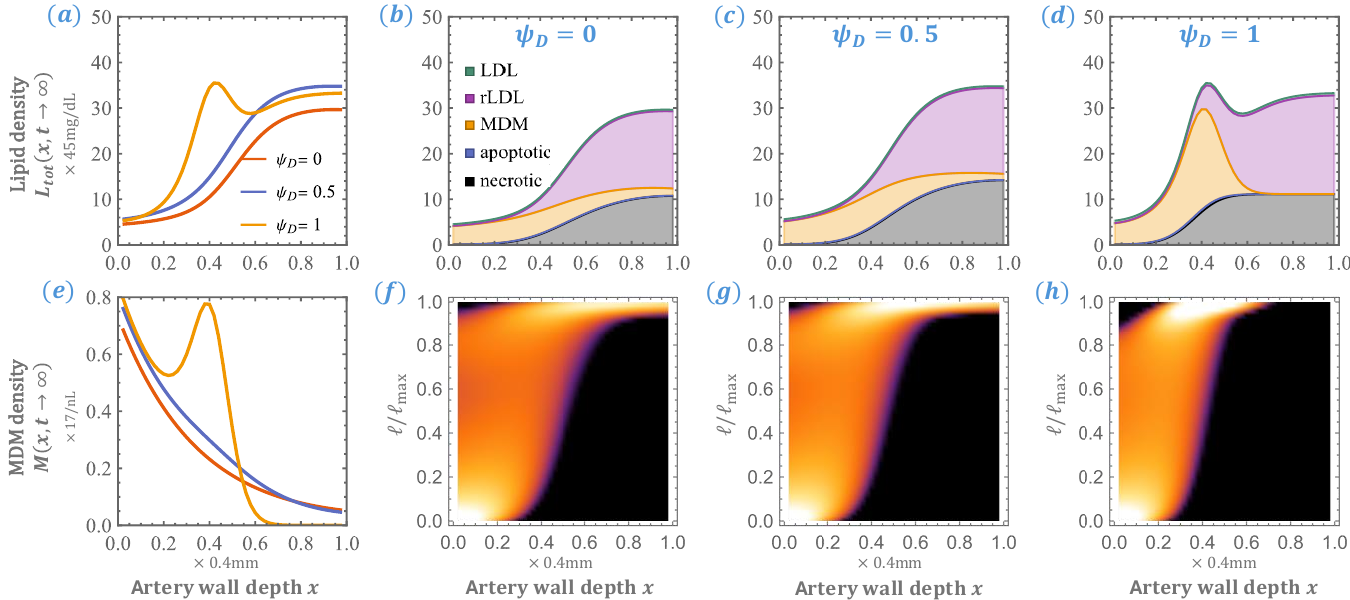}
    \caption{\textbf{Steady state lesion composition when MDM mobility decreases with lipid content: $\mathbf{\psi_D > 0, \psi_\beta = 0}$.} We compute steady state solutions for $\psi_D = 0.0, 0.5, 1.0$, with $L^\star = 4$, $H^\star = 1$. Plot (a) shows the total lipid density and (b)-(d) the densities of the lipid sub-types. Plot (e) depicts the density of MDMs and (f)-(h) the lipid-spatial MDM distribution. Note how increases in $\psi_D$ yield an accumulation of lipid and MDMs towards $x = 0$, with the formation of local maxima in the total lipid and MDM densities for $\psi_D$ = 1.}
    \label{fig: steady1_PsiD}
\end{figure}

Fig. \ref{fig: steady1_PsiD} shows typical steady state solutions for $\psi_D = 0$.0, $0.5$ and $1.0$. Plot (a) shows that increases in $\psi_D$ yield greater total lipid densities, primarily closer to the endothelium. Moreover, the lipid density attains a local maximum at $x \approx 0.4$ for $\psi_D = 1$. Comparing plots (c) and (d) to (b) shows that this increase in lipid content and peak formation reflects an accumulation of MDM lipid for $x < \frac{1}{2}$, with minimal MDM lipid for $x > \frac{1}{2}$. Plot (e) shows that the MDM density also attains a local maximum at $x \approx 0.4$ for $\psi_D = 1$, with a rapid decrease in density with $x$ to the right of the peak. Plots (f)-(h) show that MDM lipid loads are typically greater deeper in the lesion for all $\psi_D = 0.0, 0.5, 1.0$. Comparing Figs. \ref{fig: steady1_PsiD}(g, h) to Figs. \ref{fig: steady1_PsiB}(g, h) shows that there are considerably more lipid-laden MDMs deeper in the lesion when $\psi_D = 0.5, 1.0$ and $\psi_\beta = 0.0$ than when $\psi_\beta = 0.5, 1.0$ and $\psi_D = 0.0$. This is because introducing  lipid-dependent mobility, $\psi_D > 0$, promotes the retention of lipid-laden MDMs, while introducing lipid-dependent apoptosis, $\psi_\beta > 0$, promotes their rapid death. 

\begin{figure}[h]
    \centering
    \includegraphics[width=0.99\textwidth]{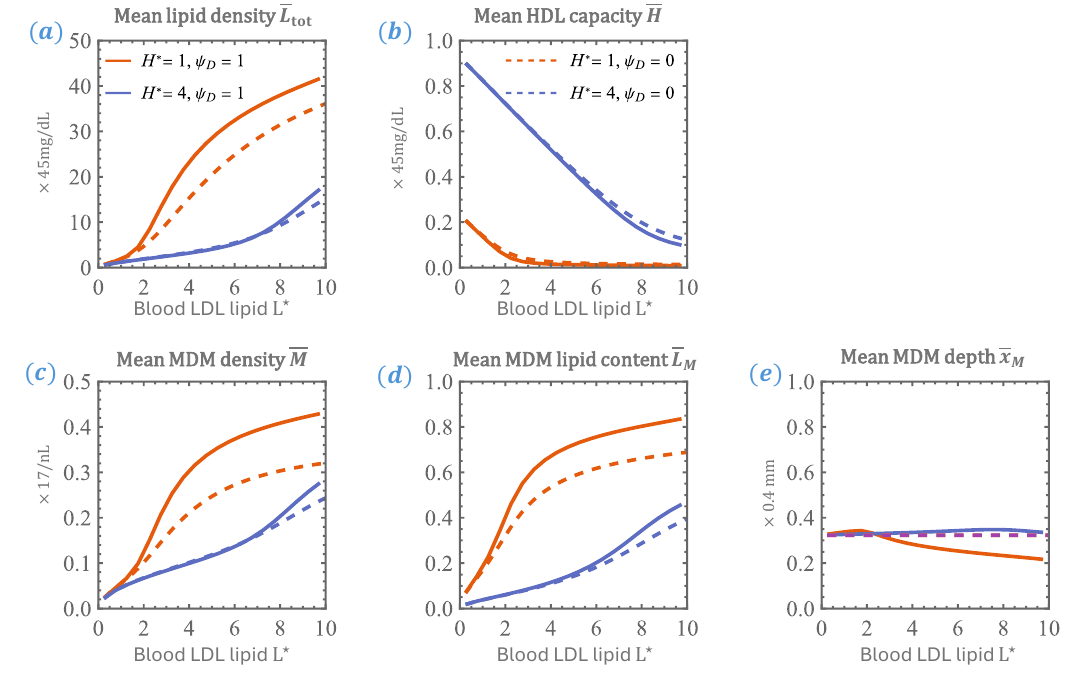}
    \caption{\textbf{Steady state summary statistics when MDM mobility decreases with lipid content (solid) vs. lipid-independent mobility (dashed).} Plots (a) and (b) show the mean lipid density and HDL capacity, defined by Eqs. \eqref{eqn: Ltot H mean}. Plots (c), (d) and (e) respectively depict the increase in mean MDM density, lipid content, and reduction in depth for $\psi_D = 1$, particularly for $H^\star  = 1$. We note again that the differences between the outputs for $\psi_D = 1$ and $\psi_D = 0$ are greater for higher values of blood LDL content, $L^\star$, and lower values of blood HDL capacity, $H^\star$. }
    \label{fig: steady2_PsiD}
\end{figure}

We compare summary statistics for the model with  lipid-dependent mobility, $\psi_D = 1$, to the lipid-independent case, $\psi_D = 0$, in Fig. \ref{fig: steady2_PsiD}. Plot (a) shows that mean total lipid density is higher for $\psi_D = 1$. This reflects how MDMs are less able to emigrate from the lesion. Indeed, by contrast to Figs. \ref{fig: steady2_PsiB}(c,d) for $\psi_\beta$, Figs. \ref{fig: steady2_PsiD}(c,d) show that MDM density and mean lipid content are greater when $\psi_D = 1$ than when the MDM kinetics are independent of lipid. The greater amount of MDM lipid for $\psi_D = 1$ gives rise to a higher efflux rate to HDL, which yields the (marginally) lower HDL capacity levels seen in plot (b). Plot (e) indicates that mean MDM depth is also decreased for $\psi_D = 1$ when $H^\star = 1$, with only minimal difference compared to the $\psi_D = 0$ case when $H^\star = 4$. In general, the differences between the plots for $\psi_D = 1$ and $\psi_D = 0$ are greater when $H^\star = 1$ and HDL capacity is exhausted (see plot (b)). The lack of HDL capacity means that lipid-laden MDMs are retained deep in the lesion since they cannot recover their mobility by offloading lipid to HDL. 

To briefly probe why $\psi_D > 0$ allows for the possibility of a local maximum in the MDM density (and hence, total lipid density), we rearrange Eq. \eqref{eqn: M nondim} at steady state, assuming only $\psi_\beta = 0$, to obtain
\begin{align}
    \frac{d^2 M}{d x^2} &= \psi_D \sum_{\ell = 0}^{\ell_\text{max}} \bigg( \frac{\ell}{\ell_\text{max}} \bigg) \frac{d^2 m_{\ell}}{d x^2} + \frac{M}{D_M} = \frac{\psi_D}{\kappa} \frac{d^2}{dx^2} (L_{M} - M) + \frac{M}{D_M} .
\end{align}
Hence $M$ may become concave, $\frac{d^2 M}{dx^2} < 0$, if the MDM ingested lipid density, $L_M - M$, is sufficiently concave:
\begin{align}
    \frac{d^2}{dx^2} (L_{M} - M) < -\frac{\kappa M}{\psi_D D_M}.
\end{align}

\subsection{Lipid-dependent mobility and apoptosis: $\psi_D, \psi_\beta > 0$} \label{sec: lipid_dependent_both}

We now suppose that both MDM mobility and apoptosis depend on lipid content: $\psi_D$, $\psi_\beta > 0$. 

\begin{figure}
    \centering
    \includegraphics[width=0.99\textwidth]{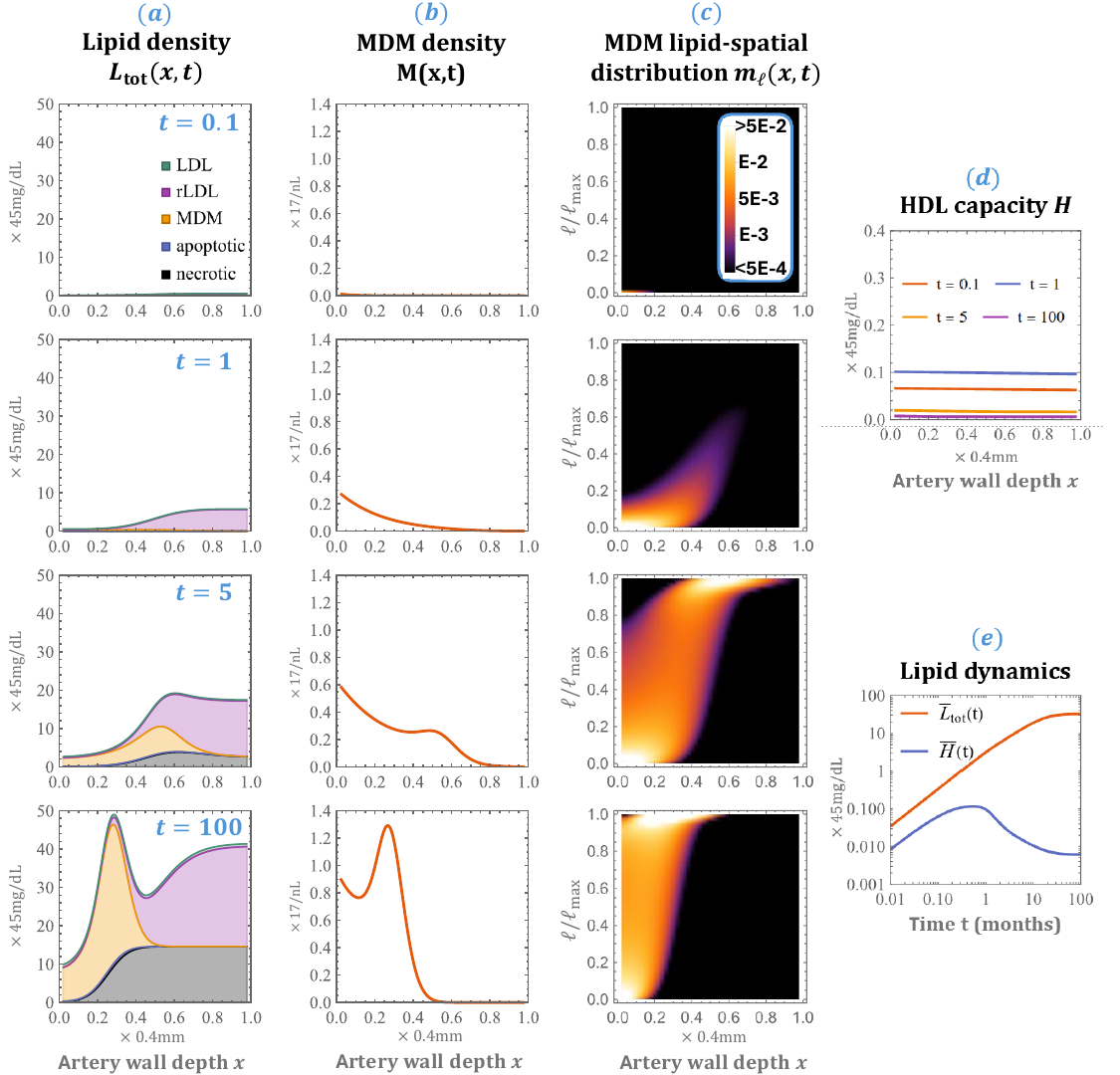}
    \caption{\textbf{Time evolution of the model lesion when MDM mobility and apoptosis depend on lipid content: $\psi_D = 1, \psi_\beta = 0.1$.} Columns (a), (b) and (c) show the lipid densities, MDM density, and MDM lipid-spatial distribution, respectively, at the times $t = 0.1$, $1$, $5$ and $100$. Plot (d) highlights the spatial uniformity of HDL capacity, as indicated by result \eqref{eqn: H asymptotic1}-\eqref{eqn: H asymptotic2}. Plot (e) shows the time evolution of mean total lipid density, $\overline{L}_\text{tot}$, and HDL capacity, $\overline{H}$, as defined in Eqs. \eqref{eqn: Ltot H mean}. Note the peak formation at later times in columns (a) and (b), consistent with the Nakashima images. We use $L^\star = 4$, $H^\star = 0.5$. }
    \label{fig: dynamics_LipidDependent}
\end{figure}

Fig. \ref{fig: dynamics_LipidDependent} shows an example solution for $\psi_D = 1.0$, $\psi_\beta = 0.1$. The model output provides an improved qualitative match to the Nakashima images (Fig. \ref{fig: Nakashima}) than the case of lipid-independent MDM kinetics (c.f. Fig. \ref{fig: dynamics_ConstantRates}). Specifically, column (a) shows that the model captures both the early accumulation of rLDL lipid deep in the lesion ($t \leq 1$), and subsequent formation of a lipid density peak towards the endothelium due to MDM lipid ($t \geq 5$). Column (b) indicates that the MDM density remains concentrated in the lesion interior for all times, with a negligible quantity of MDMs exiting the lesion at $x = 1$. Indeed, $M(x,t\geq 100) \approx 0$ for $x > 0.5$, with only a slightly larger spatial support at earlier times (e.g., $M(x, 5) \approx 0$ for $x > 0.8$). Column (c) shows that mean MDM lipid content increases with depth at all times, consistent with Fig. \ref{fig: dynamics_ConstantRates}(c), albeit with a steeper rate of increase. Figs. \ref{fig: dynamics_LipidDependent}(d,e) are similar to the case lipid-independent MDM kinetics; HDL capacity remains spatially uniform with unimodal time evolution, while mean lipid density increases monotonically with time. 

\begin{figure}
    \centering
    \includegraphics[width=0.99\textwidth]{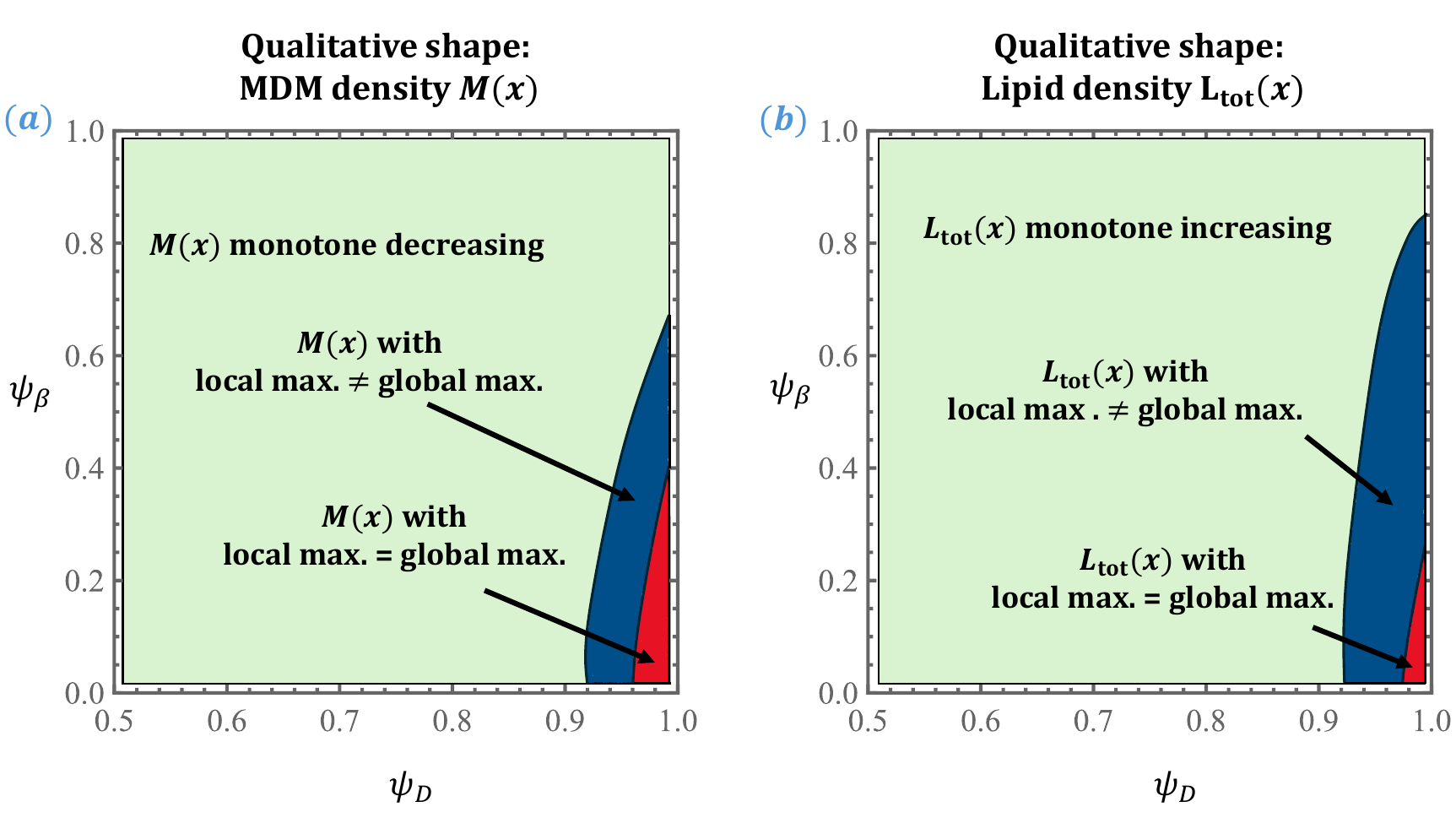}
    \caption{\textbf{The qualitative shape of the MDM $\mathbf{(a)}$ and lipid $\mathbf{(b)}$ densities as the sensitivities to lipid content of MDM diffusivity, $\psi_D$, and apoptosis, $\psi_\beta$, are varied.} The results represent steady state solutions across $(\psi_D, \psi_\beta) \in [0,1]\times[0,1]$  using a grid resolution of $0.1$ for $\psi_D < 0.9$ and $0.01$ for $\psi_D \geq 0.9$. We fix $L^\star = 4$ and $H^\star = 0.5$ and plot only $\psi_D > 0.5$ for visibility. Note how only a small neighbourhood of $(\psi_D, \psi_\beta) = (1,0)$ admits solutions for which the MDM and lipid densities attain local maxima that coincide with their respective global maxima. }
    \label{fig: psiBD_sweep}
\end{figure}


To explore the range of model behaviours as $\psi_D$ and $\psi_\beta$ vary, we compute steady state solutions across $(\psi_D, \psi_\beta) \in [0,1]\times [0,1]$ using a grid resolution of $0.1$ for $\psi_D < 0.9$ and $0.01$ for $\psi_D \geq 0.9$. The results are summarised in Fig. \ref{fig: psiBD_sweep}. We colour the $(\psi_D, \psi_\beta)$ subspace according to the qualitative behaviour of the MDM density in plot (a), and lipid density in plot (b). The results indicate that only a strikingly small region exhibits solutions with local maxima in $M(x)$ and $L_\text{tot}(x)$. In particular, peak formation in both requires large values of $\psi_D$ ($\psi_D > 0.92$) and sufficiently small values of $\psi_\beta$ ($\psi_\beta < 0.6$). Imposing the requirement that the local maxima in $M(x)$ and $L_\text{tot}(x)$ are also global maxima, as suggested by the Nakashima images, further restricts this region to a neighbourhood of $(\psi_D, \psi_\beta) = (1,0)$. This region corresponds to MDM populations for which diffusivity decreases rapidly with lipid content, and mean lifespan decreases slowly with lipid content. 

\begin{figure}
    \centering
    \includegraphics[width=0.95\textwidth]{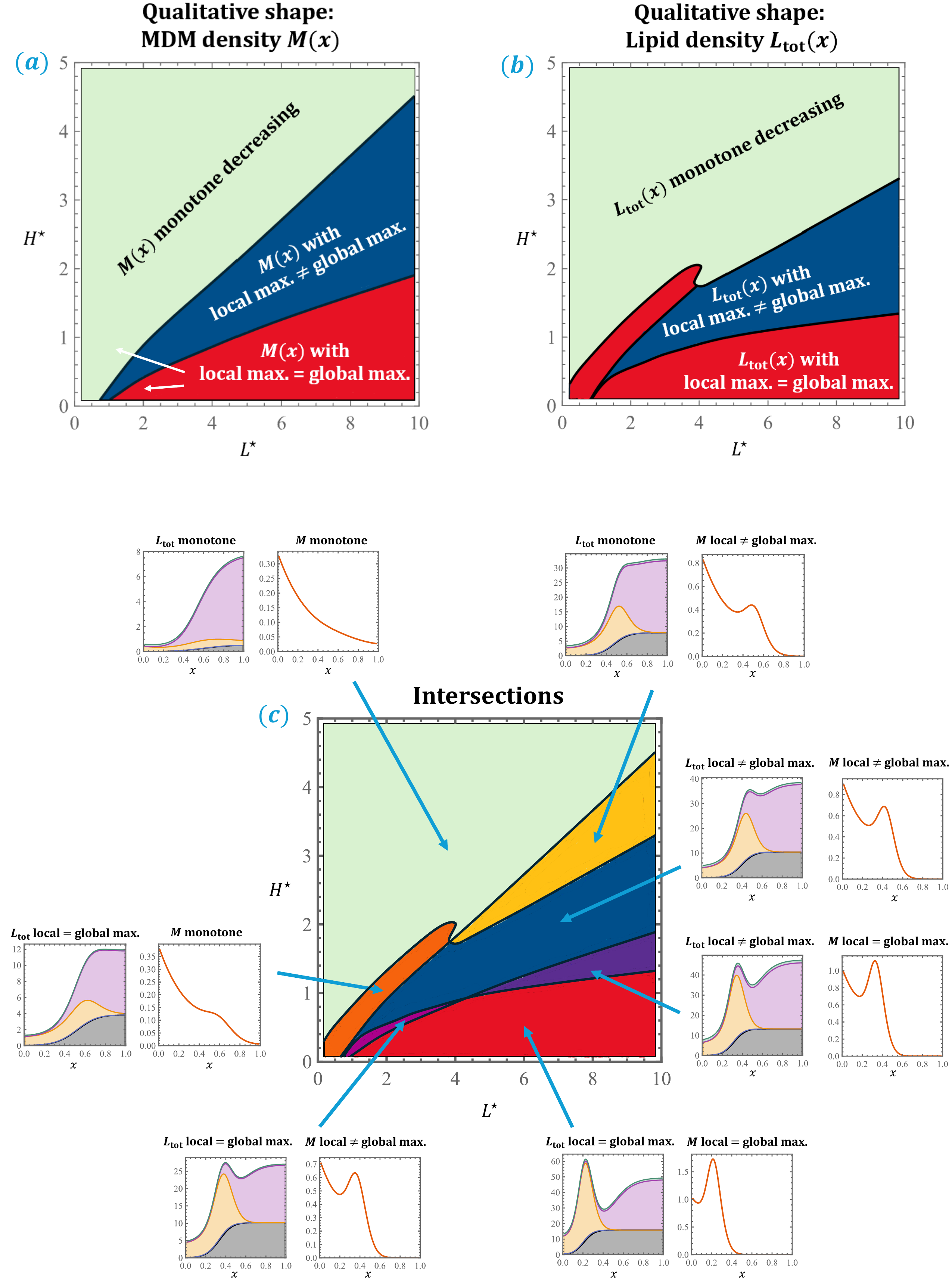}
    \caption{\textbf{The qualitative behaviour of the model with lipid content-dependent MDM mobility and apoptosis, $\psi_D$, $\psi_\beta > 0$, as blood LDL lipid, $L^\star$, and HDL capacity, $H^\star$, are varied.} We fix $\psi_D = 1$ and $\psi_\beta = 0.1$. The results represent steady state solutions across $(L^\star, H^\star) \in (0,10] \times (0,5]$ using a grid of resolution $0.2\times 0.1$. Plots (a) and (b) show the qualitative behaviour of $M(x)$ and $L_\text{tot}(x)$ at steady state, respectively, and plot (c) illustrates their intersections.
    Note how the existence of maxima in $M(x)$ and $L_\text{tot}(x)$ requires $L^\star$ to be large relative to $H^\star$. The red region in (c) depicts the range of values for which the model steady state is qualitatively consistent with the Nakashima images. }
    \label{fig: LH_sweep}
\end{figure}


We further characterise model behaviour in terms of the blood densities of LDL lipid, $L^\star$, and HDL capacity, $H^\star$, in Fig. \ref{fig: LH_sweep}. Here we fix $\psi_D = 1$ and $\psi_\beta = 0.1$, to ensure that $M(x)$ and $L_\text{tot}(x)$ exhibit local maxima that coincide with their global maxima when $L^\star = 4.5$ and $H^\star = 0.5$ (see Fig. \ref{fig: psiBD_sweep}).
The results represent steady state model output across the range of plausible values $(L^\star, H^\star) \in (0,10]\times(0,5]$. 
Plot (a) indicates that $M(x)$ exhibits a local maximum when $L^\star$ is large enough relative to $H^\star$. The local maximum coincides with the global maximum only when the difference is sufficiently large. Plot (b) shows that $L_\text{tot}(x)$ is also monotone unless $L^\star$ is sufficiently large relative to $H^\star$, although the separating curve is nonlinear. 
In contrast to the behaviour for $M(x)$, the region in which  $L_\text{tot}(x)$ is monotonic is adjacent to the region in which it is unimodal and the local and global maxima coincide. In particular, fixing $H^\star < 2$ and increasing $L^\star$ leads to the formation of a local maximum in $L_\text{tot}(x)$ that coincides with the global maximum when it exists. 
By contrast, increasing $L^\star$ for fixed $H^\star > 2$ yields a local maximum whose peak value is less than $L_\text{tot}(1)$. Plot (c) shows the intersections between plots (a) and (b). We note that there is a large region for $H^\star < 1$ and  $L^\star > 2$ where both $M(x)$ and $L_\text{tot}(x)$ attain local maxima that coincide with their global maxima, consistent with the Nakashima images. 

\begin{figure}
    \centering
    \includegraphics[width=0.99\textwidth]{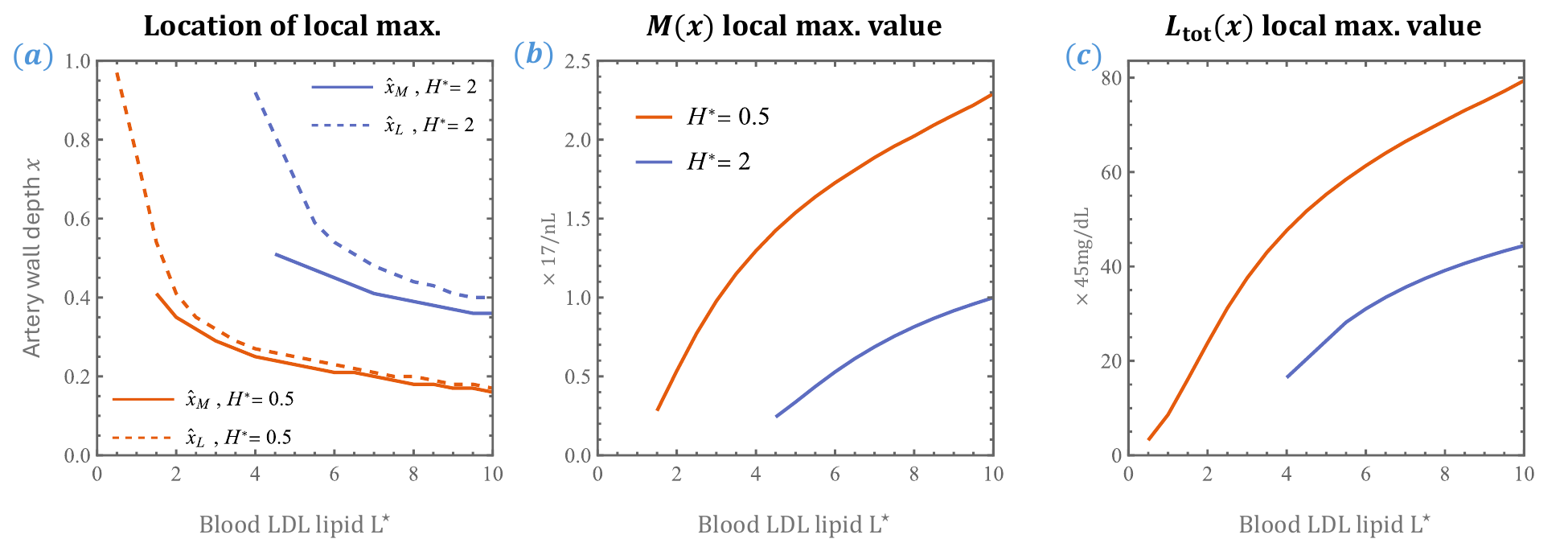}
    \caption{\textbf{Behaviour of the local maxima at steady state for the MDM and lipid densities, $M(x)$ and $L_\text{tot}(x)$, respectively. } Plot (a) shows how their peak location approximately coincides and decreases in depth for higher blood LDL levels, $L^\star$. Plots (b) and (c) show the value of the local maximum in $M(x)$ and $L_\text{tot}(x)$, respectively. Both maxima increase with $L^\star$. We fix $\psi_D = 1$, $\psi_\beta = 0.1$ and compute solutions for $H^\star = 0.5$ (blue) and $H^\star = 2$ (red).}
    \label{fig: peak_summary}
\end{figure}

Finally, we explore the behaviour of the local maxima in $M(x)$ and $L_\text{tot}(x)$ at steady state in Fig. \ref{fig: peak_summary}. We compute solutions  for $0 < L^\star \leq 10$ with $H^\star = 0.5$ (red) and $H^\star = 2$ (blue). Plot (a) shows that the location of the local maxima for $M$ (solid) and $L_\text{tot}$ (dashed) approximately coincide, with better agreement for larger $L^\star$. The location of both maxima decreases with $L^\star$, indicating that higher blood LDL levels push the local maxima towards the endothelium. Peak location is deeper for $H^\star = 2$ than for $H^\star = 0.5$, suggesting that increasing blood HDL capacity has the opposite effect, pushing the peak towards the tunica media. Plots (b) and (c) show the local maximum value of $M(x)$ and $L_\text{tot}(x)$, respectively. As expected, peak values are greater for higher blood LDL levels, $L^\star$, and lower values of HDL capacity, $H^\star$.  Overall, the results presented in Figs. \ref{fig: LH_sweep} and \ref{fig: peak_summary} show that blood levels of LDL lipid and HDL capacity strongly influence the qualitative behaviour of the lipid and MDM spatial density profiles. 

\FloatBarrier


\section{Discussion} \label{sec: discussion}

In this paper we have developed a spatially-resolved and lipid-structured model for MDM populations in early human atherosclerotic lesions. We consider a one-dimensional Cartesian geometry, where $0 \leq x \leq X$ represents depth into the artery wall. Lipid content is treated using a discrete index, $\ell = 0, 1, \dots, \ell_\text{max}$, to represent incremental changes to MDM lipid burden. The model couples the dynamics of the MDM {\color{black} densities}, $m_\ell(x,t)$, to the densities of free LDL lipid, $L_{\text{\scaleto{LDL}{3.5pt}}}(x,t)$, retained LDL lipid, $L_\text{r}(x,t)$, apoptotic lipid, $L_\text{ap}(x,t)$, necrotic lipid, $L_\text{n}(x,t)$, HDL lipid capacity, $H(x,t)$, and inflammatory mediators, $S(x,t)$.

Our model development and analysis were guided by the lipid and MDM densities presented in Nakashima \textit{et al.} \cite{nakashima2007early}, reproduced in Fig. \ref{fig: Nakashima}. Intima growth is minimal between the images, supporting our assumption of a fixed domain size: $X \approx 0.4$mm. The images exhibit two {\color{black} key} features which we aimed to qualitatively replicate in our model:
\begin{enumerate}
    \item[1.] Lipid first accumulates deep in the intima, towards the tunica media;
    \item[2.] The lipid and MDM densities both eventually peak within the tunica intima, rather than at the lumenal or medial boundaries;
\end{enumerate}
We accounted for feature 1 by assuming a spatially non-uniform LDL retention capacity, $K_\text{r}(x)$, which increases with $x$ in a step-like manner. The steepness of the step is determined by a parameter, $\theta \approx 10$, which we approximate by comparison to Fig. \ref{fig: Nakashima}(e, h). We note that it is insufficient to assume a spatially uniform LDL retention capacity, or to omit LDL retention entirely, since LDL enters the lesion from the bloodstream; the LDL spatial profile would skew towards the endothelium, contrary to Fig. \ref{fig: Nakashima}(e, h). Feature 2 indicates behaviour that should fall within the scope of our model output. Since it pertains to lipid and MDM densities at the PIT stage, we interpret this feature as a property our model should be capable of exhibiting at steady state. This is reasonable since PIT is the most advanced pre-thrombotic stage prior to fibrous cap formation. Fibrous cap formation is driven by the later accumulation of smooth muscles cells towards the endothelium; capturing these dynamics is beyond the scope of our model.  

Our model analysis in Sect. \ref{sec: results} focussed on the impact of two key parameters, $\psi_\beta, \psi_D \in [0,1]$. The values of $\psi_\beta$ and $\psi_D$ respectively determine the sensitivity of MDM lifespan and MDM mobility to lipid content. These parameters alter how the dependent variables in our model are coupled to each other. In particular, the extracellular lipid and HDL densities form a closed subsystem when coupled to the total MDM density, $M(x,t)$, and lipid content, $L_M(x,t)$, when MDM lifespan and mobility are both independent of lipid content: $\psi_\beta = \psi_D = 0$. No such closed subsystem exists if either $\psi_\beta$ or $\psi_D \neq 0$ since the PDE for $L_M(x,t)$ becomes coupled to higher order moments of the lipid distribution, $m_\ell(x,t)$, whose dynamics depend on 
successively higher order moments. We also explored the impact of the blood densities of LDL lipid and HDL capacity, $L^\star$ and $H^\star$, respectively. As both $L^\star$ and $H^\star$ depend on lifestyle choices, such as diet and exercise, it should be possible to change their values. We considered their impact across a range of plausible values: $L^\star \in (0,10)$ and $H^\star\in (0,5)$, with upper bounds corresponding to $450$mg/dL and $225$mg/dL respectively. We fixed the remaining parameters to the values given in Table \ref{tab: parameters}. Since these estimates are likely to carry high degrees of uncertainty, we focus our analysis on identifying qualitative trends rather than quantitative predictions. We discuss our findings below in relation to the key questions posed in Sect. \ref{sec: Intro}.

\begin{enumerate}
    \item[Q1.] {\color{black} \textbf{When does our model reproduce the key qualitative features of Fig. \ref{fig: Nakashima}?}}
    
The results 
presented in Sect. \ref{sec: lipid-independent} suggest that our model cannot reproduce an internal peak in the lipid or MDM densities when MDM lifespan and mobility are independent of lipid content ($\psi_\beta = \psi_D = 0$). Despite successfully capturing the early lipid accumulation deep in the intima, the lipid and MDM densities maintain monotone increasing and decreasing profiles, respectively, as the model tends to steady state. A key issue is that the MDM density at steady state $M(x)$, decouples from the lipid densities. We used this to prove that $M(x)$ decreases monotonically for all parameter values, and that the mean MDM depth is independent of $L^\star$ and $H^\star$.

The steady state results  in Sect. \ref{sec: lipid_dependent_apoptosis} indicate that if only the rate of apoptosis is lipid-dependent 
then the model is unable 
to replicate the Nakashima images. Although a closed-form analytical solution for $M(x)$ at steady state is not readily attainable when $\psi_\beta > 0 = \psi_D$, we proved that $M(x)$ is concave-up wherever it is nonzero and, hence, cannot attain an internal global maximum. 

By contrast, the steady state results 
in Sect. \ref{sec: lipid_dependent_diffusion} show that 
if only MDM mobility is lipid content-dependent then the MDM and lipid densities may 
attain internal global maxima. The steady state plots suggest that peak formation only occurs for $\psi_D$ sufficiently large.

We confirm this hypothesis in Sect. \ref{sec: lipid-independent}, where we explore model behaviour in the case where both MDM lifespan and mobility depend on lipid content: $\psi_D, \psi_\beta > 0$. Our results indicate that the formation of internal global maxima in both the lipid and MDM densities only occurs when MDM mobility is highly sensitive to lipid content, $\psi_D > 0.98$, and MDM lifespan is comparably insensitive, $\psi_\beta < 0.2$. A further parameter sweep across plausible blood densities of LDL lipid, $L^\star$, and HDL capacity, $H^\star$, indicates that the formation of internal global maxima in both the lipid and MDM densities requires that $L^\star$ is sufficiently large relative to $H^\star$. This finding is unsurprising since higher blood LDL densities promote higher MDM lipid contents, and so they amplify the impact of lipid content-dependent MDM lifespan and mobility on the lipid and MDM densities. Low levels of blood LDL lipid relative to HDL capacity give rise to few lipid-laden MDMs in the lesion, promoting dynamics qualitatively similar to the case of lipid-independent MDM kinetics. 

We note two apparent discrepancies between our model output and the Nakashima images in Fig. \ref{fig: Nakashima}. First, comparing Fig. \ref{fig: Nakashima}(i) to Fig. \ref{fig: dynamics_LipidDependent}(b) at $t = 1$, suggests that our model overestimates MDM levels at early times. Since our model predicts lesion behaviour in noiseless conditions, this discrepancy may simply be due to noise in MDM numbers at early times. 
Indeed, Fig. \ref{fig: Nakashima}(i) is not completely devoid of MDMs. Moreover, the structure of our mathematical model ensures that MDMs can only enter the lesion after LDL retention, consistent with Fig. \ref{fig: Nakashima}. Hence, this apparent difference is quantitative in nature and could be rectified by adjusting the recruitment parameter $\alpha$. Nonetheless, it would be useful to run multiple simulations using the same parameter values in an equivalent stochastic model to explore the impact of noise in MDM numbers.
Second, our model with lipid content-dependent MDM lifespan and mobility predicts that the MDM density infiltrates deeper into the lesion before retreating and forming a peak closer to the endothelium (see Fig. \ref{fig: dynamics_LipidDependent}(b) at $t = 5, 100$). This behaviour 
contrasts with the monotone increase of MDM infiltration depth with time suggested by the Nakashima images (see Fig. \ref{fig: Nakashima}(i, l, o, r)). A possible explanation is that our model shows a genuine intermediate stage of deep MDM infiltration, not captured in Fig. \ref{fig: Nakashima}. This would mean that maximal MDM depth is a poor marker for disease progression, relative to total lipid density (which monotonically increases with time in our models). Since MDM levels are low deeper in the lesion, verifying this hypothesis would require analysis of many sample lesions to eliminate noise. Another possibility is that the retreat in maximal MDM depth is an artefact of our assumption of linearity in the decrease of MDM mobility with lipid content. This hypothesis could be explored via numerical simulations with nonlinear dependence of mobility on lipid content. \\

\item[Q2.] {\color{black} \textbf{How do lipid content-dependent MDM lifespan and mobility impact lesion composition?}} \\

The inclusion of lipid content-dependent MDM lifespan and mobility has minimal impact on model behaviour at early times, relative to the lipid-independent case (compare Fig. \ref{fig: dynamics_ConstantRates} and Fig. \ref{fig: dynamics_LipidDependent} for $t \leq 1$). This is because there are few lipid-laden MDMs early in lesion development. The impact of lipid content-dependent MDM lifespan and mobility are most apparent at later times, as the model tends to steady state.

We explored the impact on the steady state solutions of allowing MDM lifespan to decrease with lipid content in Sect. \ref{sec: lipid_dependent_apoptosis}. Relative to the constant lifespan case, we saw increases in the mean total lipid density and mean HDL capacity, and decreases in mean MDM density, MDM lipid content and depth. The decrease in mean MDM density and increase in mean lipid density are consistent with the analysis of lipid content-dependent apoptosis in the spatially-homogeneous model of Watson \textit{et al.} \cite{watson2023lipid}. However, their results suggest that mean MDM lipid load is higher in the case of lipid content-dependent lifespan. This discrepancy is likely due to differences in the treatment of MDM lipid content; the model in \cite{watson2023lipid} allows for unrestricted lipid contents, while the present model imposes a maximal capacity for lipid content, $\ell_\text{max}$. In our formulation, MDMs are less likely to consume lipid if they are already lipid-laden. Therefore,  lipid-laden MDMs are more likely to die than to ingest the excess apoptotic and necrotic lipid sourced by the death of nearby MDMs, reducing mean MDM lipid content overall. 

We analysed the effect of lipid content-dependent MDM mobility on the steady state solutions in Sect. \ref{sec: lipid_dependent_diffusion}. Relative to the constant mobility case, we saw increases to mean total lipid density, MDM density and MDM lipid content, and decreases to mean HDL capacity and MDM depth. The increases in total lipid content, MDM density and MDM lipid content are consistent with the analysis of lipid content-dependent emigration in the spatially-homogeneous model of Watson \textit{et al.} \cite{watson2023lipid}. HDL capacity and spatial depth lack analogues in their model. These results are, nonetheless, intuitive on theoretical grounds. HDL capacity is lower since MDM lipid loads are greater and so MDMs offload greater amounts of lipid to HDL. MDM depth is decreased since MDMs are, on average, less mobile. \\
\item[Q3.] {\color{black} \textbf{Is MDM lipid content correlated with spatial depth?}} \\

Our results indicate that mean MDM lipid content increases with spatial depth at all times. This trend is robust to changes in the sensitivities of MDM lifespan and mobility to lipid content, $\psi_\beta$ and $\psi_D$ respectively, and to the blood densities of LDL lipid and HDL capacity, $L^\star$ and $H^\star$ respectively. These findings are unsurprising since LDL is initially retained deep in the lesion, forming a gradient of extracellular lipid that promotes higher MDM uptake rates deeper in the intima. Moreover, MDMs deeper in the intima are likely to be older (in terms of residence time) on average, than those near the endothelium; deeper MDMs have had more time to ingest lipid than nascent MDMs.

All generated plots of the MDM lipid-spatial distribution, $m_\ell(x,t)$, further suggest that the spread in MDM lipid content also decreases with depth at all times. This is also likely to be a consequence of the greater mean residence time of MDMs deeper in the intima; it is less likely to find nascent, lipid-poor MDMs deeper in the intima than MDMs which have neared an equilibrium between lipid uptake and efflux. This hypothesis could be explored more rigorously by computing an equivalent agent-based model. 
\end{enumerate}

\noindent \textbf{Future directions and conclusions. } There are many ways in which we could extend our model. For instance, the MDM population could be further structured according to phenotype, as in Chambers \textit{et al.} \cite{chambers2024blood}. The inclusion of a second structure index, $\phi$, would enable a more detailed treatment of mediator densities and MDM recruitment. A phenotype-resolved model may also be used to explore the impact of `SPM therapies' that aim to modulate MDM phenotype via exogenous application of resolving mediators \cite{back2019inflammation}. This is the topic of a future study.

The model is currently posed on a fixed spatial domain. As seen in Fig. \ref{fig: Nakashima}, this is justified between the late DIT and early PIT stages. However, we could extend the model's temporal range of validity by relaxing this assumption and including free boundaries. To extend into earlier times (e.g., to account for DIT) the endothelium could be fixed and the internal elastic lamina made free, to account for Glagov remodelling \cite{glagov1987compensatory}. To extend into later times (e.g., to account for late PIT or fibrous cap formation) the endothelium should also be made free to allow for lumenal narrowing. We may account for free boundaries by recasting the current model (or an appropriately simplified version) in a morphoelastic or multiphase framework, such as those developed by Fok \cite{fok2020finite} or Ahmed \textit{et al.} \cite{ahmed2023macrophage, ahmed2024hdl}, respectively. 

In conclusion, in this paper we have presented a new spatially-resolved and lipid-structured mathematical model for MDM populations in early human atherosclerotic lesions. Our model development and analysis is guided by the clinical observations of Nakashima \textit{et al.} \cite{nakashima2007early}. Consistent with their findings, the model predicts that lipid initially accumulates deep in the intima due to a spatially non-uniform LDL retention capacity. Our model is further capable of qualitatively reproducing their reported lipid and MDM densities provided that MDM mobility decreases sensitively with lipid content and MDM lifespan is sufficiently insensitive to lipid content. We further find that mean MDM lipid content increases with spatial depth, regardless of blood LDL and HDL content. These results develop the current understanding of macrophage heterogeneity with respect to lipid content and spatial position.


\backmatter





\bmhead{Acknowledgments}

KC acknowledges support from the Oxford Australia Scholarships Fund and Clarendon Scholars' Association. KC and HMB would like to thank the Isaac Newton Institute for Mathematical Sciences, Cambridge, for support and hospitality during the programme Mathematics of movement: an interdisciplinary approach to mutual challenges in animal ecology and cell biology, where work on this paper was undertaken. This work was supported by EPSRC grant EP/R014604/1. MRM and HMB acknowledge funding from Australian Research Council grant  DP200102071.



\section*{Declarations}

The authors declare that they have no conflict of interest.

\bibliography{sn-bibliography}


\end{document}